\pdfoutput=1
\documentclass[iop]{emulateapj}
\usepackage{apjfonts}

\newcommand{\A}{$A_{\rm o}$}

\newcommand{\muHz}{\mbox{$\mu$Hz}}
\newcommand{\teff}{$T_{\rm eff}$}
\newcommand{\teffo}{$T_{\rm eff,0}$}

\slugcomment{Accepted for publication in the Astrophysical Journal}
\accepted{April 10, 2012}

\shorttitle{Whole Earth Telescope Observations of EC14012-1446}
\shortauthors{Provencal et al.}

\begin{document}


\title{Empirical Determination of Convection Parameters in White Dwarfs I :\\ Whole Earth Telescope Observations of EC14012-1446
\footnotemark[$\dagger$]}
\footnotetext[$\dagger$]{Based on observations obtained at the Southern Astrophysical Research (SOAR) telescope, which is a joint project of the 
Minist\'{e}rio da Ci\^{e}ncia, Tecnologia, e Inova\c{c}\~{a}o (MCTI) da Rep\'{u}blica Federativa do Brasil, 
the U.S. National Optical Astronomy Observatory (NOAO), the University of North Carolina at Chapel Hill (UNC), and Michigan State University (MSU).}


\author
{J.~L.~Provencal\altaffilmark{1,2},
M.~H.~Montgomery\altaffilmark{2,3},
A.~Kanaan\altaffilmark{4},
S.~E.~Thompson\altaffilmark{1,2,5},
J.~Dalessio\altaffilmark{1,2},
H.~L.~Shipman\altaffilmark{1,2},
D.~Childers\altaffilmark{6,2},
J.~C.~Clemens\altaffilmark{7},
R.~Rosen\altaffilmark{8},
P.~Henrique\altaffilmark{4},
A.~Bischoff-Kim\altaffilmark{9},
W.~Strickland\altaffilmark{10},
D.~Chandler\altaffilmark{10},
B.~Walter\altaffilmark{10},
T.~K.~Watson\altaffilmark{11},
B.~Castanheira\altaffilmark{12},
S.~Wang\altaffilmark{3},
G.~Handler\altaffilmark{12},
M.~Wood\altaffilmark{13},
S.~Vennes\altaffilmark{13},
P.~Nemeth\altaffilmark{13},
S.~O.~Kepler\altaffilmark{14},
M.~Reed\altaffilmark{15},
A.~Nitta\altaffilmark{16},
S.~J.~Kleinman\altaffilmark{16},
T.~Brown\altaffilmark{17},
S.~-L.~Kim\altaffilmark{18},
D.~Sullivan\altaffilmark{19},
Wen-Ping~Chen\altaffilmark{20},
M.~Yang\altaffilmark{20},
Chia-You~Shih\altaffilmark{20},
X.~J.~Jiang\altaffilmark{21},
A.~V.~Sergeev\altaffilmark{22},
A.~Maksim\altaffilmark{22},
R.~Janulis\altaffilmark{23},
K.~S.~Baliyan\altaffilmark{24},
H.~O.~Vats\altaffilmark{24},
S.~Zola\altaffilmark{25,26},
A.~Baran\altaffilmark{25},
M.~Winiarski\altaffilmark{25,26},
W.~Ogloza\altaffilmark{25,26},
M.~Paparo\altaffilmark{27},
Z.~Bognar\altaffilmark{27},
P.~Papics\altaffilmark{27},
D.~Kilkenny\altaffilmark{29},
R.~Sefako\altaffilmark{28},
D.~Buckley\altaffilmark{28},
N.~Loaring\altaffilmark{28},
A.~Kniazev\altaffilmark{28},
R.~Silvotti\altaffilmark{30},
S.~Galleti\altaffilmark{30}
T.~Nagel\altaffilmark{31},
G.~Vauclair\altaffilmark{32},
N.~Dolez\altaffilmark{33},
J.~R.~Fremy\altaffilmark{33},
J.~Perez\altaffilmark{34},
J.~M.~Almenara\altaffilmark{34},
L.~Fraga\altaffilmark{14}
} 
\altaffiltext{1}{University of Delaware, Department of Physics and Astronomy
Newark, DE 19716; jlp@udel.edu}
\altaffiltext{2}{Delaware Asteroseismic Research Center, Mt. Cuba Observatory,
Greenville, DE 19807}
\altaffiltext{3}{Department of Astronomy, University of Texas, Austin, TX 78712;
mikemon@rocky.as.utexas.edu}
\altaffiltext{4}{Departamento de F\'{\i}sica Universidade Federal de Santa
Catarina, C.P. 476, 88040-900, Florian{\'o}polis, SC, Brazil; ankanaan@gmail.com}
\altaffiltext{5}{SETI Institute, NASA Ames Research Center, Moffett Field, CA 94035}
\altaffiltext{6}{Department of Math and Science, Delaware County Community
College, 901 S. Media Rd, Media, PA 19063; dpc@udel.edu}
\altaffiltext{7}{University of North Carolina 288 Phillips Hall, Chapel Hill, NC
27599; clemens@physics.unc.edu}
\altaffiltext{8}{NRAO Green Bank, WV 24944; rachel.rosen@gmail.com}
\altaffiltext{9}{Georgia College and State University, Department of Chemistry
and Physics, Milledgeville, GA 31061;agnes.kim@gcsu.edu}
\altaffiltext{10}{Meyer Observatory and Central Texas Astronomical Society, 209
Paintbrush, Waco, TX 76705; chandler@vvm.com}
\altaffiltext{11}{Southwestern University, Georgetown, TX;
tkw@sousthwestern.edu}
\altaffiltext{12}{Institut f\"ur Astronomie Universit\"at Wien,
T\"urkenschanzstrasse 17, 1180, Austria;  gerald@camk.edu.pl}
\altaffiltext{13}{Florida Institute of Technology, Dept. of Physics \& Space
Sciences, 150 W Univ. Blvd, Melbourne, FL 3290; wood@fit.edu}
\altaffiltext{14}{Instituto de F\'{\i}isica UFRGS, C.P. 10501, 91501-970 Porto
Alegre,RS, Brazil; kepler@if.ufrgs.br}
\altaffiltext{15}{Missouri State University and Baker Observatory, 901S.
National, Springfield, MO 65897; MikeReed@missouristate.edu}
\altaffiltext{16}{Gemini Observatory, Northern Operations Center, 670 North
A'ohoku Place, Hilo, HI 96720; atsuko.nittakleinman@gmail.com}
\altaffiltext{17}{Las Cumbres Observatory Global Telescope Network, Inc.  6740
Cortona Dr. Suite 102 Santa Barbara CA 93117; tbrown@lcogt.com}
\altaffiltext{18}{Korea Astronomy and Space Science Institute, Daejeon 305-348,
Korea; slkim@kasi.re.kr}
\altaffiltext{19}{School of Chemical \& Physical Sciences, Victoria University
of Wellington, P.O. Box 600, Wellington, New Zealand;denis.sullivan@vuw.ac.nz}
\altaffiltext{20}{Lulin Observatory, National Central University, Taiwan;
wchen@astro.ncu.edu.tw}
\altaffiltext{21}{National Astronomical Observatories, Academy of Sciences,
Beijing 100012, PR China; xjjiang@bao.ac.cn}
\altaffiltext{22}{Ukrainian National Academy of Sciences, Main Astronomical
Observatory, Golosiiv, Kiev 022 252650; sergeev@terskol.com }
\altaffiltext{23}{Institute of Theoretical Physics and Astronomy, Vilnius
University, Vilnius, Lithuania; jr@itpa.lt}
\altaffiltext{24}{Physical Research Laboratory, Ahmedabad 380009, India}
\altaffiltext{25}{Mount Suhora Observatory, Cracow Pedagogical University, Ul.
Podchorazych 2, 30-084 Krakow, Poland; zola@astro1.as.ap.krakow.pl}
\altaffiltext{26}{Astronomical Observatory, Jagiellonian University, ul. Orla
171, 30-244 Cracow, Poland}
\altaffiltext{27}{Konkoly Observatory, P.O. Box 67, H-1525 Budapest XII,
Hungary; paparo@konkoly.hu }
\altaffiltext{28}{South African Astronomical Observatory, PO Box 9, Observatory
7935, South Africa}
\altaffiltext{29}{Department of Physics, University of the Western Cape, Private
Bag X17, Belville 7535, South Africa}
\altaffiltext{30}{INAF-Osservatorio Astronomico di Capodimonte, via Moiariello
16, 80131 Napoli, Italy}
\altaffiltext{31}{Institut f\"ur Astronomie und Astrophysik, Universi\"at
T\"ubingen, Sand 1, 72076 T\"ubingen, Germany; nagel@astro.uni-tuebingen.de }
\altaffiltext{32}{Laboratoire d'Astrophysique de Toulouse-Tarbes, Universit\'e
de Toulouse, CNRS, 14 avenue Edouard Belin, F314000 Toulouse, France; gerardv@srvdec.obs-mip.fr }
\altaffiltext{33}{Observatoire de Paris, LESIA, 92195 Meudon, France}
\altaffiltext{34}{Instituto de Astrofisica de Canarias, 38200 La Laguna,
Tenerife, Spain}


\begin{abstract}
  We report on analysis of 308.3 hrs of high speed photometry
  targeting the pulsating DA white dwarf EC14012-1446. The data were
  acquired with the Whole Earth Telescope (WET) during the 2008
  international observing run XCOV26. The Fourier transform of the
  light curve contains 19 independent frequencies and numerous
  combination frequencies. The dominant peaks are 1633.907, 1887.404,
  and 2504.897 \muHz. Our analysis of the combination amplitudes
  reveals that the parent frequencies are consistent with modes of
  spherical degree $l$=1.  The combination amplitudes also provide $m$
  identifications for the largest amplitude parent frequencies. Our
  seismology analysis, which includes 2004--2007 archival data,
  confirms these identifications, provides constraints on additional
  frequencies, and finds an average period spacing of 41 s. Building
  on this foundation, we present nonlinear fits to high
  signal-to-noise light curves from the SOAR 4.1m, McDonald 2.1m, and
  KPNO 2m telescopes. The fits indicate a time-averaged convective
  response timescale of $\rm{\tau_0=99.4\pm17}$~s, a temperature
  exponent $N=85\pm6.2$ and an inclination angle of
  $\theta_i=32.9\pm3.2^\circ$.  We present our current empirical map
  of the convective response timescale across the DA instability
  strip.
\end{abstract}


\keywords{stars:white dwarfs ---stars:oscillations ---asteroseismology: general
---white dwarfs: individual(EC14012-1446) ---stars:evolution
\vspace*{-1em}}

\section{Introduction}\label{sec:intro}

Stellar seismology, also known as \emph{asteroseismology}, provides us
with a unique tool for probing the interiors of stars, allowing us to
study fundamental problems in stellar evolution such as energy
transport, thermodynamics, and magnetism. White dwarfs, the
evolutionary endpoint of most stars, are particularly important
targets for asteroseismology. They are structurally simple: an
electron degenerate carbon/oxygen core surrounded by thin
non-degenerate layers of hydrogen and helium. DA white dwarfs
represent $\sim$ 80 percent of all white dwarfs \citep{Eisenstein06a},
and they have a nearly pure layer of hydrogen on top of a layer of
helium. DB white dwarfs lack this hydrogen layer, having a layer of
nearly pure helium overlying a carbon/oxygen core. Lacking substantial
nuclear reactions, white dwarfs simply cool as they age, passing
through specific temperature ranges (the DBV and DAV instability
strips) within which they pulsate. These pulsators are otherwise
normal objects, so what we learn about their structure can be applied
to the entire population of stellar remnants, and further applied to
our understanding of their main sequence progenitors.

In this paper, we focus on combining asteroseismology of the DAV
EC14012-1456 with nonlinear analysis of its light curve to provide an
empirical description of its convection zone. Convection is one of the
largest remaining sources of theoretical uncertainty in astrophysical
modeling. Convection is traditionally treated through mixing length
theory \citep[MLT,][]{Bohm-Vitense58}. MLT is a simple, local, time
independent description first applied to stellar modeling by
\citet{Biermann32}.  It describes the motions of ``average''
convective cells with a mean size $l=\alpha \,H_p$, where $H_p$ is the
local pressure scale height and $\rm{\alpha}$ is an adjustable free
parameter. MLT is clearly incomplete; we know turbulent flows are
complex and there is no reason why $\rm{\alpha}$ should remain
constant throughout the convection zone of a single star, and
certainly not for stars of different masses, chemical compositions, or
evolutionary phases. As an example for white dwarfs,
\citet{Bergeron95} and \citet{Tremblay2010} show that model parameters
such as flux, line profiles, energy distribution, color indices, and
equivalent widths are extremely sensitive to the assumed MLT
parameterization. \citet{Bergeron95} find systematic uncertainties
ranging from 25 percent for effective temperatures to 11 percent for
mass and radius. The use of MLT to treat convective energy transport
in white dwarf atmospheres thus represents a significant source of
physical uncertainty in calculations of their atmospheric
structure. We rely on these models to supply the information about
white dwarf interiors, masses, and temperatures needed to calibrate
white dwarf cooling sequences. This in turn produces detailed age
estimates for white dwarfs \citep{Ruiz01} and an estimate of the age
of the Galactic disk \citep{Winget87,Harris06}.  An observational test
of MLT that leads to an improved description of convection is an
important goal that will have implications beyond the study of white
dwarfs.

\begin{figure}[t]
 \epsscale{1.0}
\centering{
\includegraphics[width=1.00\columnwidth]{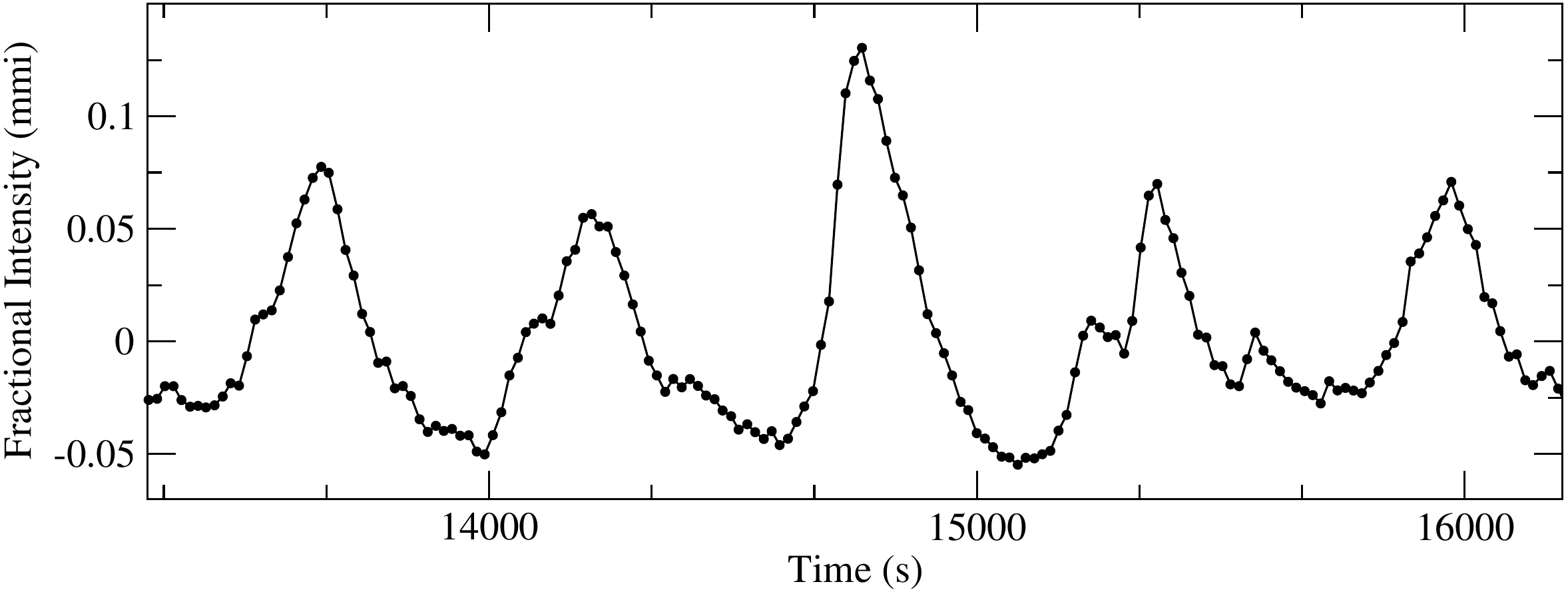}
}
\caption{Portion of a high signal-to-noise SOAR light curve of
  EC14012-1446, showing the narrow peaks and broader valleys
  indicative of convective mixing. The error bars for each point are
  indicated by the size of the points.  The dominant pulsation period
  is $\approx 600$~s.(1 mmi $\approx$1 mmag)}
\label{soarcurve}
\end{figure}

\citet{Brickhill92a} was the first to realize that a pulsating white
dwarf's photospheric flux is modulated relative to the flux entering
at the bottom of the convection zone by an amount that depends on the
convection zone's thickness.  The local convection zone depth is a
function of the local effective temperature, and this varies during a
pulsation cycle. The result is a distortion of the observed light
curve, with narrow peaks and wider valleys (Figure~\ref{soarcurve}).
Convective light curve fitting exploits these nonlinearities to
recover the thermal response timescale of the convection zone.
Mathematical details of this technique can be found in
\citet{Montgomery05b} and \citet{Montgomery10}. For our purposes,
applying convective light curve fitting to a target star requires
three ingredients: 1) a pulsator with a nonlinear light curve, 2)
precise knowledge of the star's pulsation frequencies and ($l,m$)
values, and 3) high signal-to-noise light curves for use in the actual
fitting process.

Asteroseismology provides the tools to identify white dwarf pulsation
($l,m$) values. White dwarfs are g-mode pulsators, and each pulsation
mode can be described by a spherical harmonic of degree $l$, radial
overtone $k$, and azimuthal number $m$, where $m$ takes
integer values between $-l$ and $l$. Given a sufficient sample of
excited pulsation modes, we can match the observed frequencies with
theoretical models. An important diagnostic for g-mode pulsators is the 
mean period spacing between modes of the same ($l,m$) but consecutive 
radial overtone $k$ (e.g., the $k$ and $k+1$ modes).  The mean period 
spacing depends mainly on stellar mass. Deviations of individual spacings 
from this mean value provide information on the thickness of the hydrogen and/or
helium layers and on the chemical profile of the core
\citep{Montgomery09}. We note here that the actual value of the radial
overtone $k$ can not be determined observationally, but must be
inferred from theoretical models.

\begin{figure*}[t]
\epsscale{1.0}
\centering{
\includegraphics[width=0.80\textwidth]{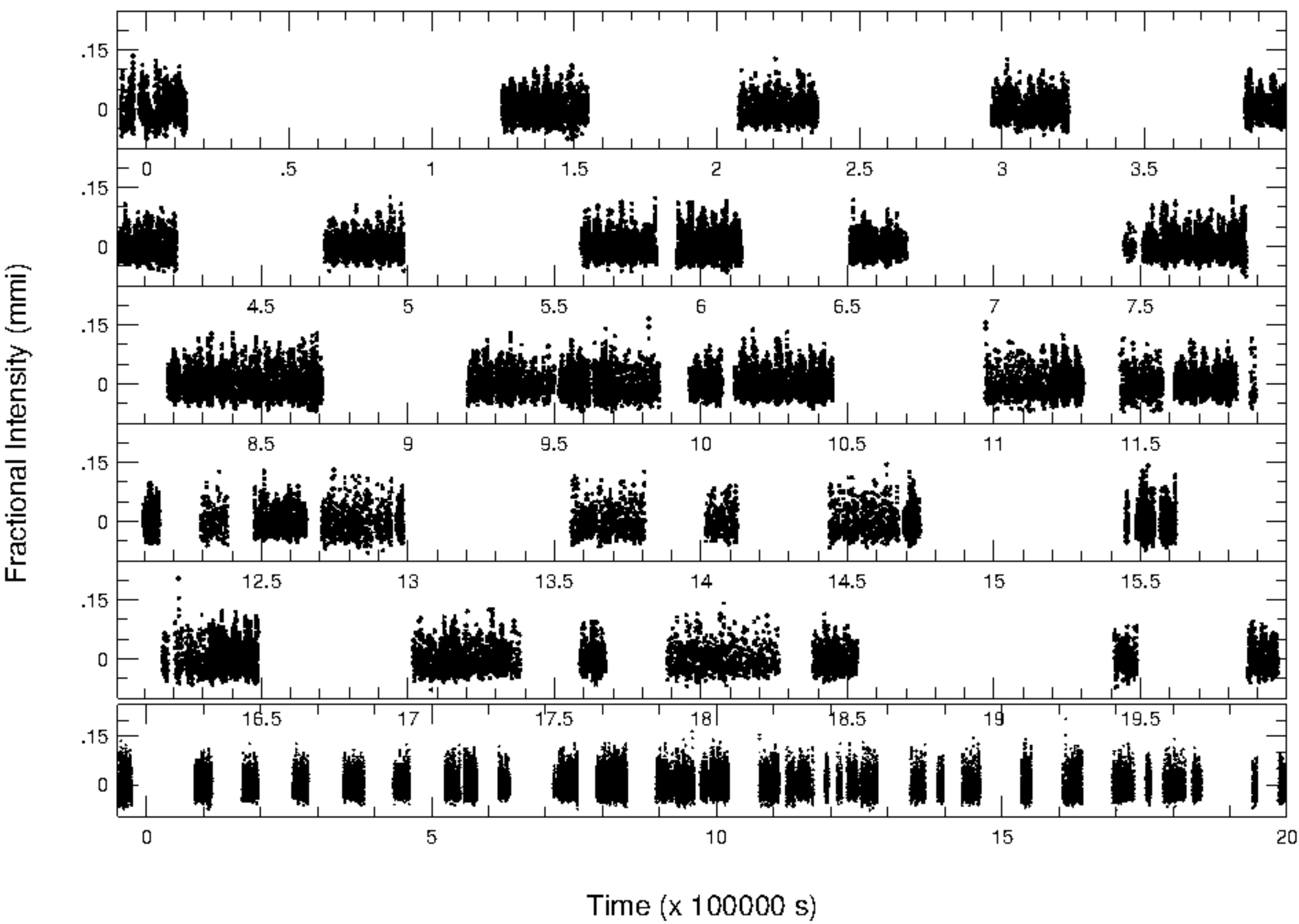}
}
\caption{The final light curve of EC14012-1446 from XCOV26, representing the combination
of all data listed in Table~\ref{tab:journal}.  The top five panels 
cover 400000 s (4.63 days) each.  The bottom panel displays the entire light curve, with time 
in days. The dominant pulsation period is $\approx$600 s.
  \label{xcovcurve}}
\end{figure*}

A second diagnostic is given by the presence of multiplets. The
multiplet components have the same $(k,l)$ and are further described
by the azimuthal index $m$, which takes integer values between $-l$
and $l$.  To first order, the frequency difference relative to the
$m=0$ component of the multiplet is given by
$\delta\nu_{klm}=-m\,\Omega\,(1-C_{kl})$, where $\Omega$ is the rotation
period and $C_{kl}$ is a coefficient that depends on the pulsation
eigenfunctions evaluated in the nonrotating case. In the high-$k$
asymptotic limit for $g$-modes, $C_{kl} \sim 1/\ell (\ell+1)$,
although models predict it to vary by $\approx 10$ percent over the
range of observed periods in EC14012. Multiplet structure is a strong
indication of a mode's $l$ value.  We expect a triplet for $l=1$, a
quintuplet for $l=2$, and so on. The observed frequency differences
(splittings) are a measure of the stellar rotation rate as sampled by
a given mode.  Deviations from equal frequency splitting within a
single multiplet and changes in splittings from one multiplet to the
next reveal information about differential rotation and magnetic field
strength.

Applying convective light curve fitting to a wide sample of pulsating
white dwarfs provides an empirical map of how the convective response
timescale varies as a function of effective temperature, and this can
be compared with theoretical models, both MLT and hydrodynamic. The
Whole Earth Telescope (WET) is engaged in a long term project to
provide such a description of convection across the hydrogen
atmosphere DAV (\teff\ $\approx$ 11100--12200K) and the helium
atmosphere DBV (\teff\ $\approx$ 22000--29000K) instability strips. We
present here our results for the DAV pulsator EC14012-1446. Our goals
are three-fold: determine accurate frequency and ($l,m$)
identifications, obtain several nights of high quality light curves,
and apply convective light curve fitting to obtain EC14012-1446's
convective response timescale.  In the following, we present the
resulting data set and our analysis of the Fourier transforms, discuss
the identified frequencies and ($l,m$) values, perform nonlinear light
curve fits, and present the convective parameters for EC14012-1446.

\section{Observations and Reductions}

EC14012-1446 (WD1401-147, B=15.67), is a high amplitude, multiperiodic
DAV pulsator discovered by \citet{Stobie95} and observed extensively
by \citet{Handler08}. Our XCOV26 observations span March 25 to April
29 2008, achieving 80 percent coverage during the central five days of the
run (Figure~\ref{xcovcurve}). Twenty-seven telescopes distributed
around the globe participated in XCOV26, fifteen of which contributed
a total of 71 runs (Table~\ref{tab:journal}) on EC14012-1446. The
observations were obtained with different CCD photometers, each with
distinct effective bandpasses. We minimize the bandpass issues by
using CCDs with similar detectors where possible and employing a red
cut-off filter (BG40 or S8612) to normalize wavelength response and
reduce extinction effects.

Standard procedure for a WET run calls for observers to transfer raw
images and calibration files to WET headquarters for analysis at the
end of each night. CCD data reduction follows the steps outlined in
\citet{Provencal09}. We corrected each image for bias and thermal
noise, and normalized by the flat-field.  Aperture photometry using
the IRAF photometry pipeline described by \citet{Kanaan02} was
performed on each image, utilizing a range of aperture sizes for the
target and selected comparison stars. We used the WQED pipeline
\citep{wqed} to examine each nightly light curve for photometric
quality, remove outlying points, divide by a suitable comparison star
and correct for differential extinction. Our observational technique
is therefore not sensitive to oscillations longer than a few hours.
The result is light curves with amplitude variations represented as
fractional intensity (mmi). The unit is a linear representation of the
fractional intensity of modulation (1 mmi $\approx$1 mmag). We present
our Fourier transforms (FT) in units of modulation amplitude (1~mma =
$1\times10^{-3} {\rm ma} = 0.1\% = 1$~ppt).

The final reduction step combines the individual light curves to
produce the complete light curve for EC14012-1446. We assume
EC14012-1446 oscillates about a mean light level. This assumption
allows us to carefully assess overlapping segments from different
telescopes and identify and correct any vertical offsets. As discussed
in detail in \citet{Provencal09}, we find no significant differences
between the noise levels of amplitude spectra using: 1) the
combination of all light curves including overlapping segments from
different telescopes, 2) the combination of light curves where we
retain only higher signal to noise observations in overlapping
segments and 3) combining all light curves using data weighted by
telescope aperture.

\begin{deluxetable*}{rrrcc} 
\tablecolumns{5}
\tablewidth{0pc}
\tabletypesize{\small}
\tablecaption{Journal of XCOV26 Observations EC14012-1446}
\tablehead{
\colhead{Run Name} & \colhead{Telescope} & \colhead{Detector} & \colhead{Date} & \colhead{Length} \\
\colhead{} & \colhead{(m)} & \colhead{} &\colhead{} & \colhead{(hrs)} 
}
\startdata
 hawa20080325-09& Hawaii 2.2  & E2V ccd47-10 & 2008-03-25 & 6.2 \\
 saao20080326-21& SAAO 1.0  & UCT CCD & 2008-03-26 & 5.9 \\
 tene20080327-20 & Tenerife 0.8  & TK1024 & 2008-03-27 & 3.0 \\
 saao20080327-20 & SAAO 1.0  & UCT CCD & 2008-03-27 & 7.5 \\
 saao20080328-20 & SAAO 1.0  & UCT CCD & 2008-03-28 & 7.3 \\
 saao20080329-20 & SAAO 1.0  & UCT CCD & 2008-03-29 & 7.2 \\
 tene20080330-03 & Tenerife 0.8  & TK1024 & 2008-03-30 & 2.9 \\
 saao20080330-20 & SAAO 1.0  & UTC CCD & 2008-03-30 & 7.5 \\
 saao20080331-20 & SAAO 1.0  & UTC CCD & 2008-03-31 & 7.1 \\
 mcdo20080401-05 & McDonald 2.1  & E2V ccd57-10 & 2008-04-01 &
 \phantom{1}6.1\tablenotemark{*} \\
 saao20080401-20 & SAAO 1.0  & UTC CCD & 2008-04-01 & 5.4 \\
 loia20080403-00 & Loia 1.52  & EEV 1300x1340B & 2008-04-03 & 1.1 \\
 ctio20080403-01 & CTIO 0.9  & $\rm{Tek2K_3}$ & 2008-04-03 & 7.5 \\
 soar20080403-02 & SOAR 4.1 & 2Kx4K MIT/Lincoln Lab CCD & 2008-04-03 &
 \phantom{1}7.6\tablenotemark{*} \\
 sara20080403-05 & SARA 1.0  & CCD & 2008-04-03 & 6.3 \\
 mcdo20080403-08 & McDonald 2.1  & E2V ccd57-10 & 2008-04-03 & 3.0 \\
 saao20080403-20 & SAAO 1.0  & UCT CCD & 2008-04-03 & 6.8 \\
 salt20080403-21 & SALT 10.0 & CCD & 2008-04-03 & 1.0 \\
 soar20080404-01 & SOAR 4.1 & CCD & 2008-04-04 & \phantom{1}7.7\tablenotemark{*}\\
 salt20080404-02 & SALT 10.0 & E2V 44-82  & 2008-04-04 & 0.7 \\
 ctio20080404-02 & CTIO 0.9 & $\rm{Tek2K_3}$ & 2008-04-04 & 7.0 \\
 mcdo20080404-05 & McDonald 2.1 & E2V ccd57-10 & 2008-04-04 &
 \phantom{1}5.4\tablenotemark{*} \\
 sara20080404-10 & SARA 1.0 & CCD & 2008-04-04 & 0.6 \\
 saao20080404-21 & SAAO 1.0 & UTC CCD & 2008-04-04 & 6.4 \\
 salt20080404-21 & SALT 10.0 & E2V 44-82 &2008-04-04 & 1.0 \\
 soar20080405-01 & SOAR 4.1 & 2Kx4K MIT/Lincoln Lab CCD & 2008-04-05 &  \phantom{1}4.7\tablenotemark{*} \\
 salt20080405-02 & SALT 10.0 & E2V 44-82 & 2008-04-05 & 1.0 \\
 ctio20080405-03 & CTIO 0.9 & $\rm{Tek2K_3}$ & 2008-04-05 & 6.3 \\
 mcdo20080405-09 & McDonald 2.1 & E2V ccd57-10 & 2008-04-05 & 1.7 \\
 mtjo20080405-11 & Mt. John 1.0 & E2V ccd57-10 & 2008-04-05 & 3.3 \\
 boao20080405-14 & BOAO 1.8 & SITe SI424AB CCD & 2008-04-05 & 4.4 \\
 salt20080405-21 & SALT 10.0 & E2V 44-82& 2008-04-05 & 1.0 \\
 saao20080405-22 & SAAO 1.0 & UTC CCD & 2008-04-05 & 3.1 \\
 salt20080406-01 & SALT 10.0 & E2V 44-82 & 2008-04-06 & 1.0 \\
 soar20080406-02 & SOAR 4.1 &2Kx4K MIT/Lincoln Lab CCD & 2008-04-06 &
 \phantom{1}7.3\tablenotemark{*} \\
 mcdo20080406-09 & McDonald 2.1 & E2V ccd57-10 & 2008-04-06 & 2.0 \\ 
 ctio20080407-02 & CTIO 0.9 & $\rm{Tek2K_3}$ & 2008-04-06 & 6.2 \\
 mcdo20080407-08 & McDonald 2.1 & E2V ccd57-10 & 2008-04-07 & 3.1 \\
 boao20080407-15 & BOAO 1.8 & SITe SI424AB CCD & 2008-04-07 & 4.0 \\
 saao20080407-20 & SAAO 1.0 & UTC CCD & 2008-04-07 & 5.9 \\
 salt20080407-20 & SALT 10.0 & E2V 44-82 & 2008-04-07 & 1.0 \\
 salt20080408-01 & SALT 10.0 & E2V 44-82 & 2008-04-08 & 1.0 \\
 ctio20080408-02 & CTIO 0.9 & $\rm{Tek2K_3}$ & 2008-04-08 & 1.8 \\
 mcdo20080408-09 & McDonald 2.1 & E2V ccd57-10 & 2008-04-08 & 1.5 \\
 luli20080408-14 & Lulin 1.0 & E2V CCD36-40 & 2008-04-08 & 2.6 \\
 saao20080408-20 & SAAO 1.0 & UTC CCD & 2008-04-08 & 4.9 \\
 ctio20080409-02 & CTIO 0.9 & $\rm{Tek2K_3}$ & 2008-04-09 & 6.6 \\
 mcdo20080409-09 & McDonald 2.1 & E2V ccd57-10 & 2008-04-09 & 0.6 \\
 ctio20080410-02 & CTIO 0.9 & $\rm{Tek2K_3}$ & 2008-04-10 & 6.9 \\
 boao20080410-14 & BOAO 1.8 & SITe SI424AB CCD & 2008-04-10 & 3.0 \\
 ctio20080411-02 & CTIO 0.9 & $\rm{Tek2K_3}$ & 2008-04-11 & 6.6 \\
 mcdo20080411-09 & McDonald 2.1 & E2V ccd57-10 & 2008-04-11 & 1.5 \\
 mcdo20080412-06 & McDonald 2.1 & E2V ccd57-10 & 2008-04-12 & 0.3 \\
 kpno20080412-07 & KPNO 2.0 & E2V ccd47-10 & 2008-04-12 & 3.7 \\
 ctio20080413-02 & CTIO 0.9 & $\rm{Tek2K_3}$ & 2008-04-13 & 4.4 \\
 mcdo20080413-06 & McDonald 2.1 & E2V ccd57-10 & 2008-04-13 & 4.7 \\
 ctio20080414-02 & CTIO 0.9 & $\rm{Tek2K_3}$ & 2008-04-14 & 7.1 \\
 kpno20080414-04 & KPNO 2.0 & E2V ccd47-10 & 2008-04-14 & 2.4 \\
 lash20080414-08 & Las Cumbres 2.0 &E2V CCD42-40  & 2008-04-14 & 3.4 \\
 mcdo20080414-09 & McDonald 2.1 & E2V ccd57-10 & 2008-04-14 & 1.7 \\
 luli20080414-17 & Lulin 1.0 & E2V CCD36-40 & 2008-04-14 & 2.4 \\
 ctio20080415-02 & CTIO 0.9 & $\rm{Tek2K_3}$ & 2008-04-15 & 6.6 \\
 lash20080415-09 & Las Cumbres 2.0 & E2V CCD42-40  & 2008-04-15 & 3.8 \\
 luli20080415-16 & Lulin 1.0 & E2V ccd36-40 & 2008-04-15 & 4.3 \\
 mabu20080415-20 & Mt. Abu 1.2 & TEK CCD & 2008-04-15 & 1.3 \\
 mabu20080416-20 & Mt. Abu 1.2 & TEK CCD & 2008-04-16 & 2.2 \\
 kpno20080417-09 & KPNO 2.0 & E2V ccd47-10  & 2008-04-17 & 2.8 \\
 kpno20080418-07 & KPNO 2.0 & E2V ccd47-10 & 2008-04-18 & 3.9 \\
 braz20080426-04 & LNA 1.6 & WI106 CCD & 2008-04-26 & 2.1 \\
 braz20080427-05 & LNA 1.6 & WI106 CCD & 2008-04-27 & 1.9 \\
 braz20080428-04 & LNA 1.6 & WI106 CCD & 2008-04-28 & 1.9 \\
 braz20080429-05 & LNA 1.6 & WI106 CCD & 2008-04-29 & 1.2 
\enddata
\tablenotetext{*}{High signal to noise light curves used for light curve fitting. All other observations were used to obtain accurate frequency information.}

\label{tab:journal}
\end{deluxetable*}

\begin{figure*}[p]
  \centering{
    \includegraphics[width=0.82\textwidth]{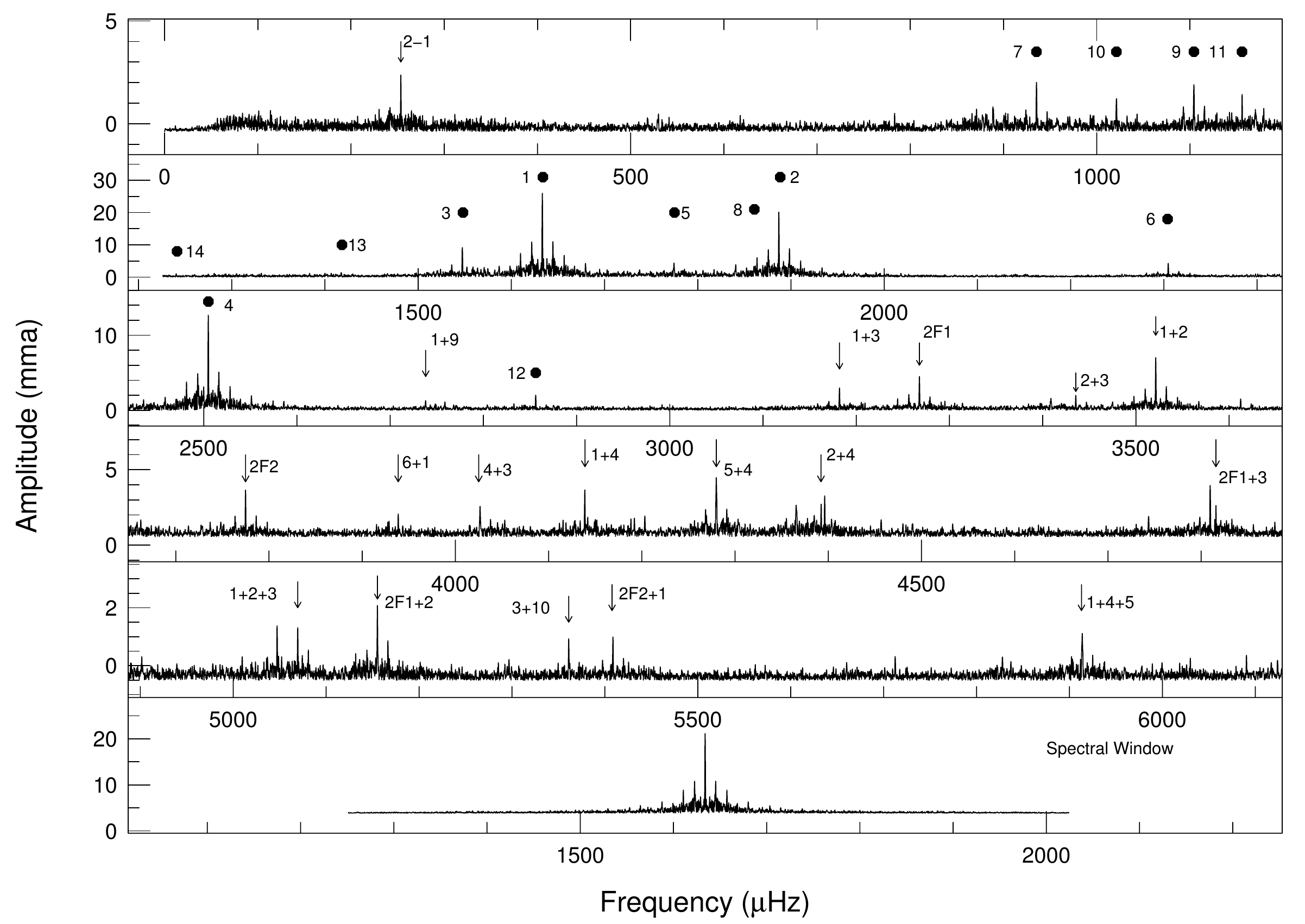}
}
\caption{Fourier Transform of the 2008 EC14012-1446 photometry
  observations (note vertical scale in each panel). Solid dots label
  14 modes containing 19 independent frequencies.  Arrows indicate a
  sample of combination frequencies (for example F1+F2). Unlabeled
  peaks are second order (for example F1+F2) and third order (for
  example F1+F2+F3) combinations.  The spectral window is plotted in
  the last panel.  Tables~\ref{tab:freq} and \ref{tab:comb} list exact
  frequency identifications.\label{xcov26dft}}
\end{figure*}

\begin{figure*}[p]
  \centering{
\includegraphics[width=0.8\textwidth]{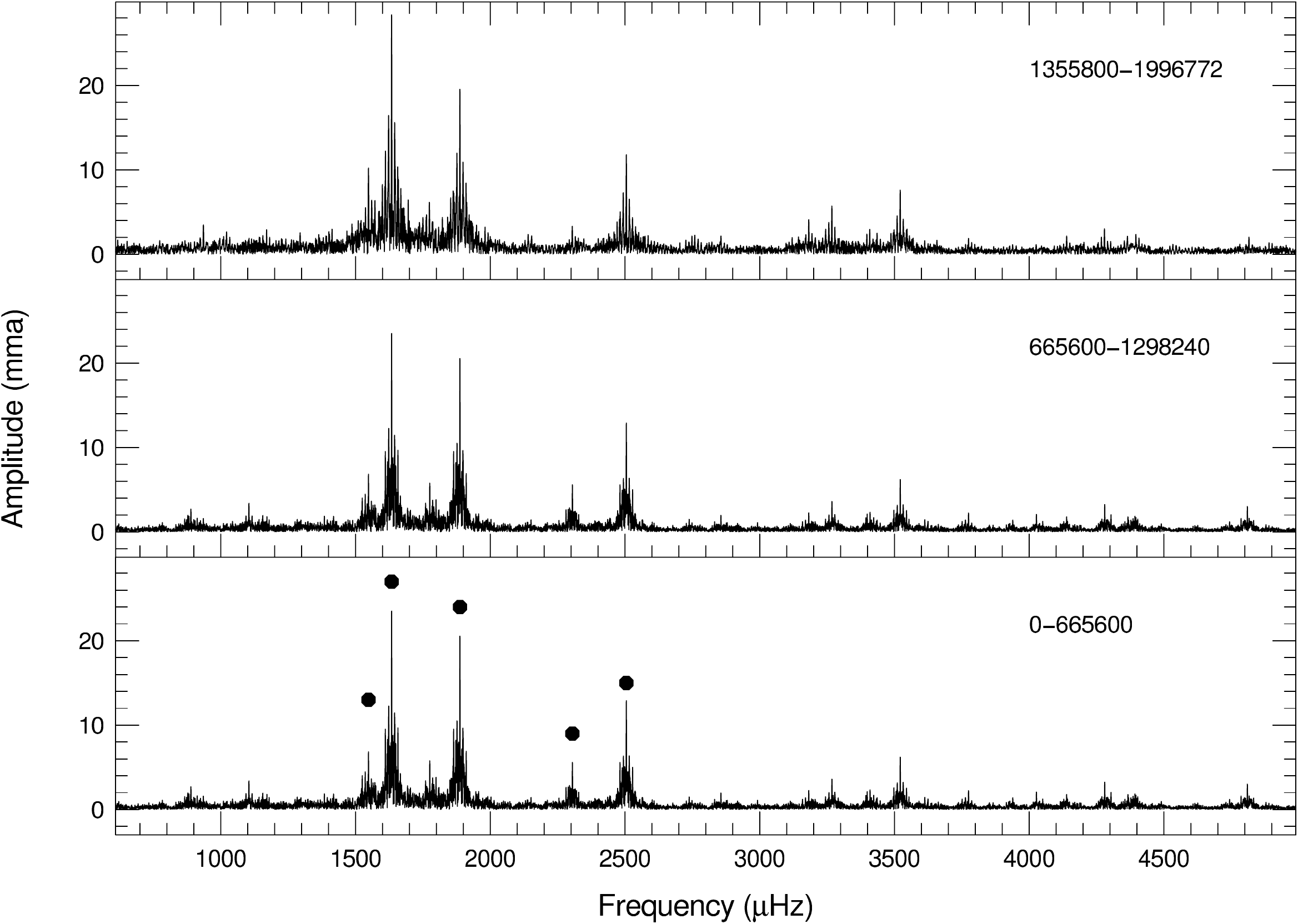}
}
\caption{FTs of the EC14012-1446 data set subdivided into three chunks
  of $\approx$185 hr each. The labels in the right of each panel give
  the time segment covered by each chunk. The black points in the
  bottom panel identify the five largest amplitude modes. Some changes
  in each FT can be explained as differences in the window structure
  and effective resolution for each chunk. In particular, the final
  chunk has the least data coverage.  We do find evidence of amplitude
  and/or frequency variations during the run. All peaks with
  frequencies higher than $\approx 3000$~\muHz\ are combination
  peaks.\label{chunks}}
\end{figure*}

The final XCOV26 light curve contains 308.3 h of high speed
photometry. Our coverage is not complete, and this incompleteness
produces spectral leakage in the amplitude spectrum. To quantify this,
we sampled a single sinusoid using exactly the same times as the
original data. The resulting amplitude spectrum, or ``spectral
window'', is the pattern produced in the XCOV26 Fourier Transform by a
single frequency. The Fourier transform (FT) and spectral window of
the complete light curve are given in Figure~\ref{xcov26dft}.

\begin{figure}[t]
  \centering{
\includegraphics[width=0.95\columnwidth]{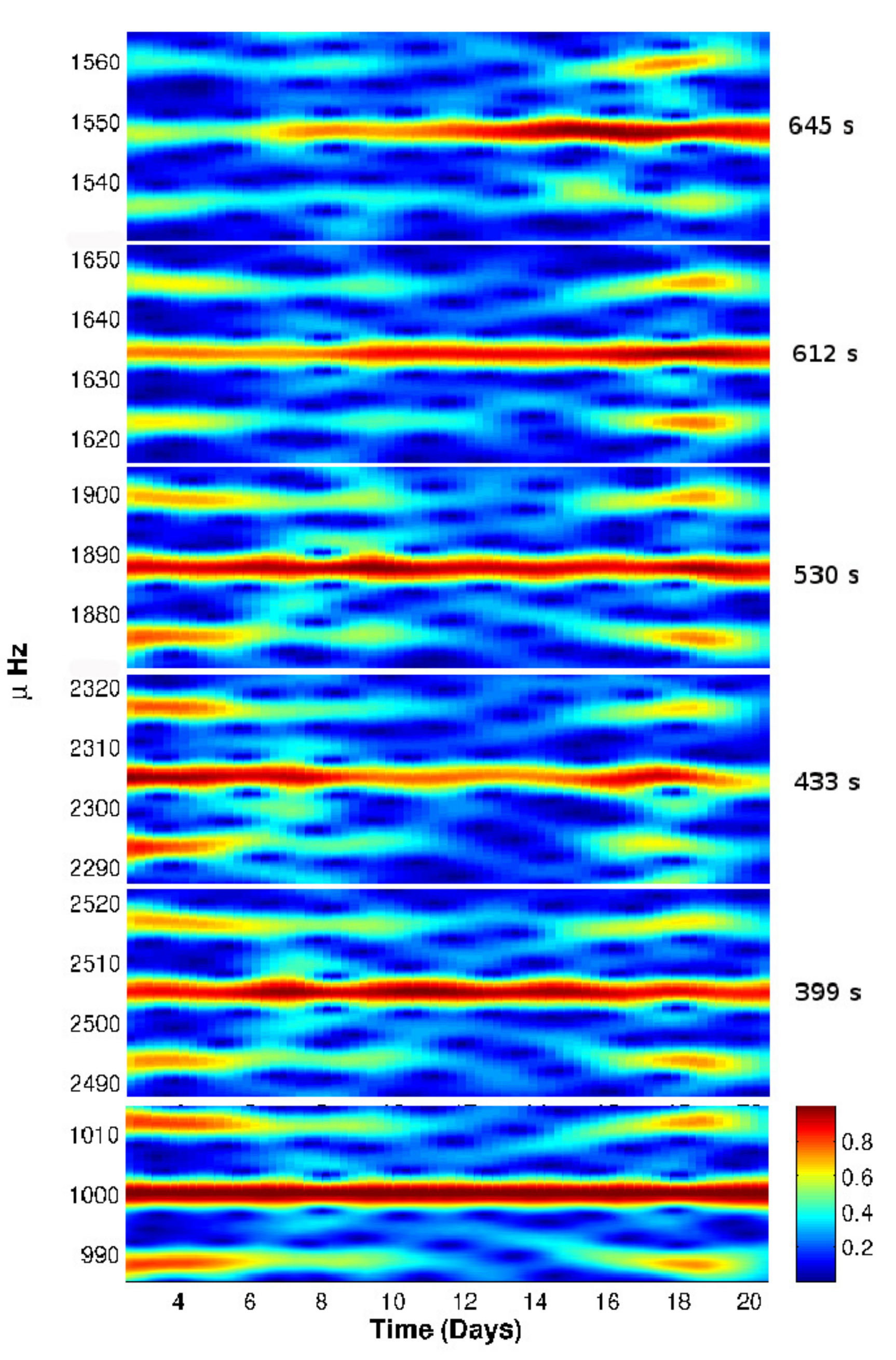}
}
\caption{Spectrogram of the five largest amplitude modes in
  EC14012-1446's FT, arranged in order of decreasing period. The
  bottom panel gives the time-varying spectral window for a sample
  frequency at 1000~\muHz\ (the pattern given by a single sinusoid
  sampled at exactly the same times as the data). Each panel is
  normalized to an amplitude of one. We find a slow amplitude increase
  in the 1633.907 \muHz\ (612 s) and 1548.146 \muHz\ (645 s) peaks. In
  addition, the 2308 \muHz\ (433 s) peak shows a decrease in amplitude
  during the middle of the run.
  \label{ec11allspgmcolor}
}
\end{figure}

\section{Frequency Identification}
\subsection{Stability}\label{sec:stab}

Following \citet{Provencal09}, we begin by examining the stability of
EC14012-1446's pulsation amplitudes and frequencies throughout the
run.  Amplitude and/or frequency variations produce artifacts in FTs,
and greatly complicate the identification of intrinsic pulsation
frequencies \citep{Kepler03}. We divide the data set into three
chunks, each spanning $\approx$185 h (7.8 d). The FT of each
chunk is given in Figure~\ref{chunks}. Several of the dominant
frequencies are consistent to within measurement error, but we do find
evidence of amplitude variations in a number of frequencies.

\begin{figure}[t]
 \centering{
\includegraphics[width=1.0\columnwidth]{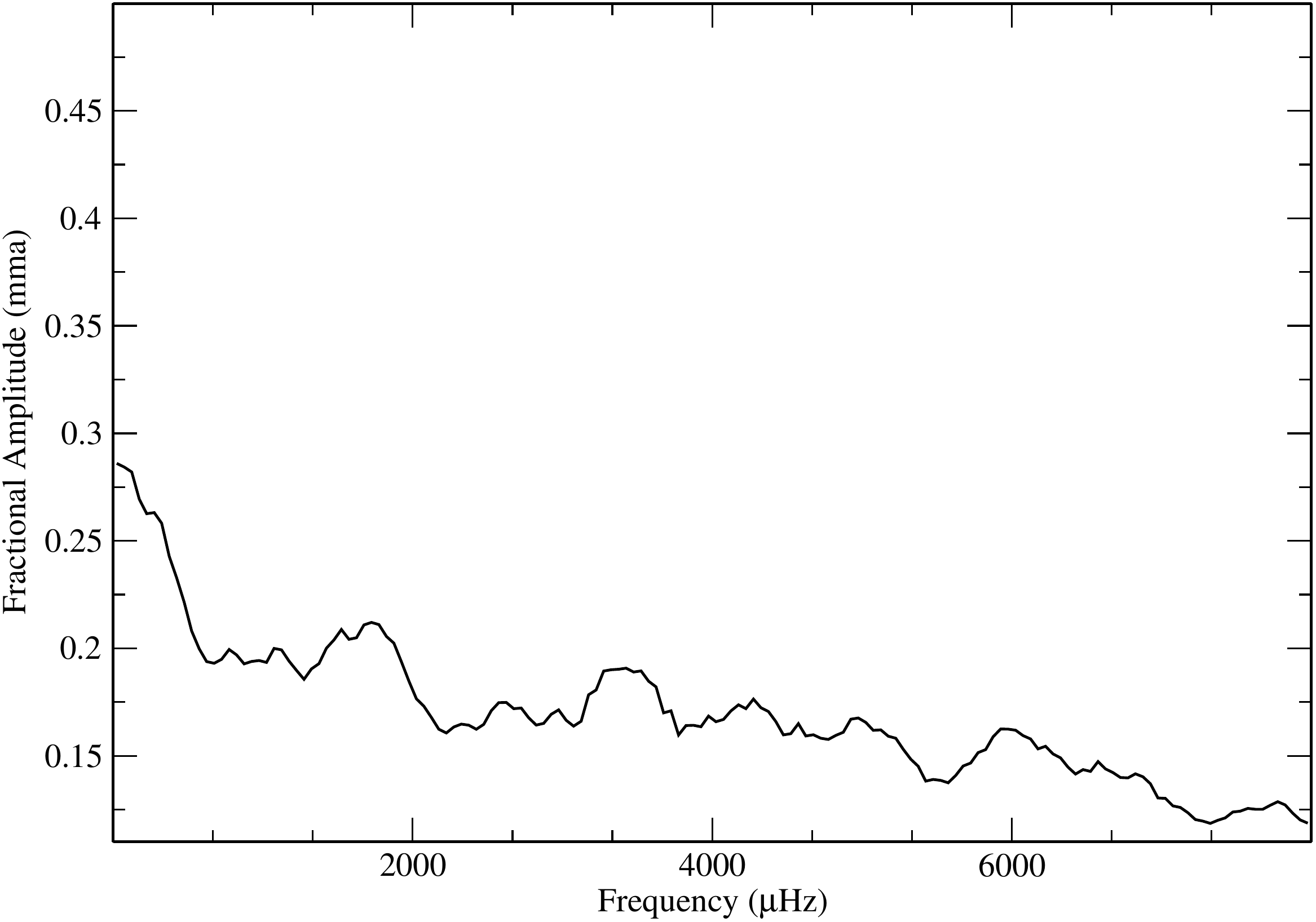}
}
\caption{The mean noise as a function of frequency for XCOV26.  The
  data set was prewhitened by the 62 largest amplitude frequencies.
  Our noise is frequency dependent, but is near 0.2 mma.  This is a
  conservative estimate, as we have probably not prewhitened by all
  the combination frequencies.}
\vspace*{0.2em}
\label{noise}
\end{figure}

We calculated spectrograms for the five largest amplitude frequencies
to further explore the nature of these amplitude variations 
(Figure~\ref{ec11allspgmcolor}). A spectrogram 
quantifies the behavior of frequencies and amplitudes as a function
of time. Our spectrograms are generated by dividing the total light
curve into multiple 5 day segments, each of which overlap by 4.95
days. The FT of each segment is a measurement of frequencies and
amplitudes centered on a specific time. Each panel in 
Figure~\ref{ec11allspgmcolor} is an amalgam,
where each segment FT corresponds to a vertical line. The x axis is
time, the y axis is frequency, and amplitude is represented by
color and is normalized to an amplitude of 1. 
The bottom panel in Figure~\ref{ec11allspgmcolor} gives the pattern
generated by a single sinusoid sampled with exactly the same times as the segment light
curves (a time-dependent spectral window). The results show a
slow 18$\sigma$ increase in the amplitude of the dominant 1633.907 \muHz\
(612~s) peak, from $\approx$ 22 to $\approx$ 31 mma, over the course
of the run. We also find a similar increase in the 1548.146 \muHz\ (645 s)
peak. The 2308 \muHz\ (433 s) peak shows a 25 percent decrease in amplitude
during the middle of the run and a large apparent decrease at the end of the run. 
The amplitudes of remaining two frequencies are consistent to within 3$\sigma$.

\subsection{The XCOV26 Fourier Transform}

Armed with amplitude stability information for the five largest peaks, we
are ready to take a careful look at EC14012-1446's XCOV26 FT.  We use
{\sl Period04} \citep{Lenz05} for Fourier analysis and nonlinear least
squares fitting to select the statistically significant peaks in the
XCOV26 FT. As detailed in \citet{Provencal09}, we adopt the criterion
that a peak must have an amplitude at least four times greater than the
mean noise level in the given frequency range.  We define ``noise''
as the frequency-dependent mean amplitude after prewhitening by the
dominant frequencies. This is a conservative estimate, as it is
impossible to ensure that all of the ``real'' frequencies are removed
when calculating the noise level. This is certainly true for
EC14012-1446, where the peaks above $\approx$ 3000 \muHz\ are mainly
combination frequencies. Figure~\ref{noise} displays the mean amplitude,
specified as the square root of the simple mean power using a boxcar 
of 100 \muHz, after prewhitening by 62 frequencies, as a function of frequency.
Our noise is somewhat frequency dependent, but is near 0.2 mma.   

To confirm our uncertainty estimates, we calculated Monte Carlo
simulations using the routine provided in {\sl Period04}. This routine
generates a set of light curves using the original times, the fitted
frequencies and amplitudes, and added Gaussian noise. A least squares
fit is performed on each simulated light curve, with the distribution
of fit parameters giving the uncertainties.  Our Monte Carlo results
are consistent with our mean amplitude noise estimates of $\approx$0.2 mma.  

\begin{figure}[t]
  \centering{
\includegraphics[width=1.0\columnwidth]{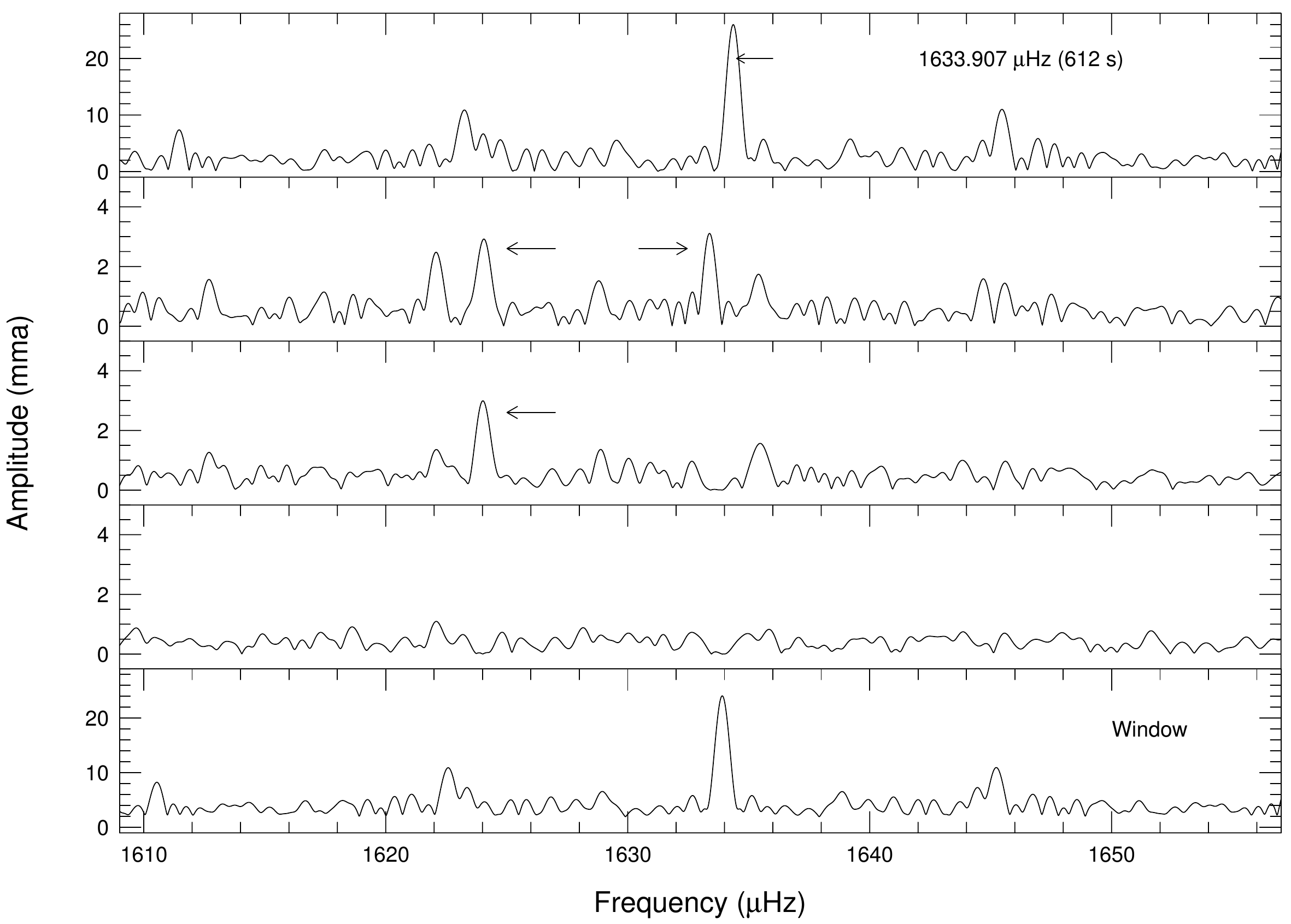}
}
\caption{Prewhitening of the dominant 1633.907 \muHz\ mode in the 2008
  FT. We begin with the removal of the largest amplitude resolved peak
  (top panel), a careful comparison of the residuals in the next panel
  with the spectral window (last panel), and the subsequent removal of
  additional frequencies. The fourth panel shows the residuals after
  simultaneously removing 1633.907, 16233.450, and 1624.017\muHz. The 1633.450 \muHz\
  component (right arrow in second panel) is result of amplitude changes.\label{w1633} }
\end{figure}

Our frequency selection procedure involves identifying the largest
amplitude resolved peak in the FT, fitting a sinusoid with that
frequency to the data set, subtracting the fit from the light curve,
recomputing the FT, examining the residuals, and repeating the process
until no significant power remains. This technique, known as
prewhitening, must be employed with an abundance of caution,
especially since we are aware of amplitude and/or frequency modulation
in our data set. The modulation will create artifacts that masquerade
as additional frequencies. To illustrate, let us examine the region of
dominant power at 1633.907 \muHz\ (Figure~\ref{w1633}). Comparison of
the original FT (top panel) with the spectral window (bottom panel)
demonstrates that most of the signal is concentrated at 1633.907
\muHz. We fit a sinusoid to the data to determine frequency,
amplitude, and phase, and subtract the result from the original light
curve. The second panel of Figure~\ref{w1633} shows the prewhitened
FT. Careful examination reveals two residual peaks (at 1633.450 and
1624.015\muHz\, identified with arrows) that are clearly not
components of the spectral window. A third peak at 1622 \muHz\ is part
of a window pattern associated with 1633.450 \muHz, but the alias
pattern is asymmetric.  Panel 3 shows the results of prewhitening by a
simultaneous fit of 1633.907 and 1633.450 \muHz.  Both frequencies are
removed, and 1622 \muHz\ is diminished, leaving 1624.015 \muHz.  We
next subtract a simultaneous fit of 1633.907, 1633.450, and 1624.015
\muHz, with the results displayed in panel 4 of Figure~\ref{w1633}. No
significant power remains. However, a red flag is raised: the
separation between 1633.907 and 1633.450 \muHz\ is 0.457 \muHz, the
inverse of which is the run length. The 1633.450~\muHz\ peak is a
manifestation of the amplitude changes observed in
Figure~\ref{ec11allspgmcolor} and is not included in our final
frequency list.

\begin{figure}[t]
  \centering{
\includegraphics[width=1.0\columnwidth]{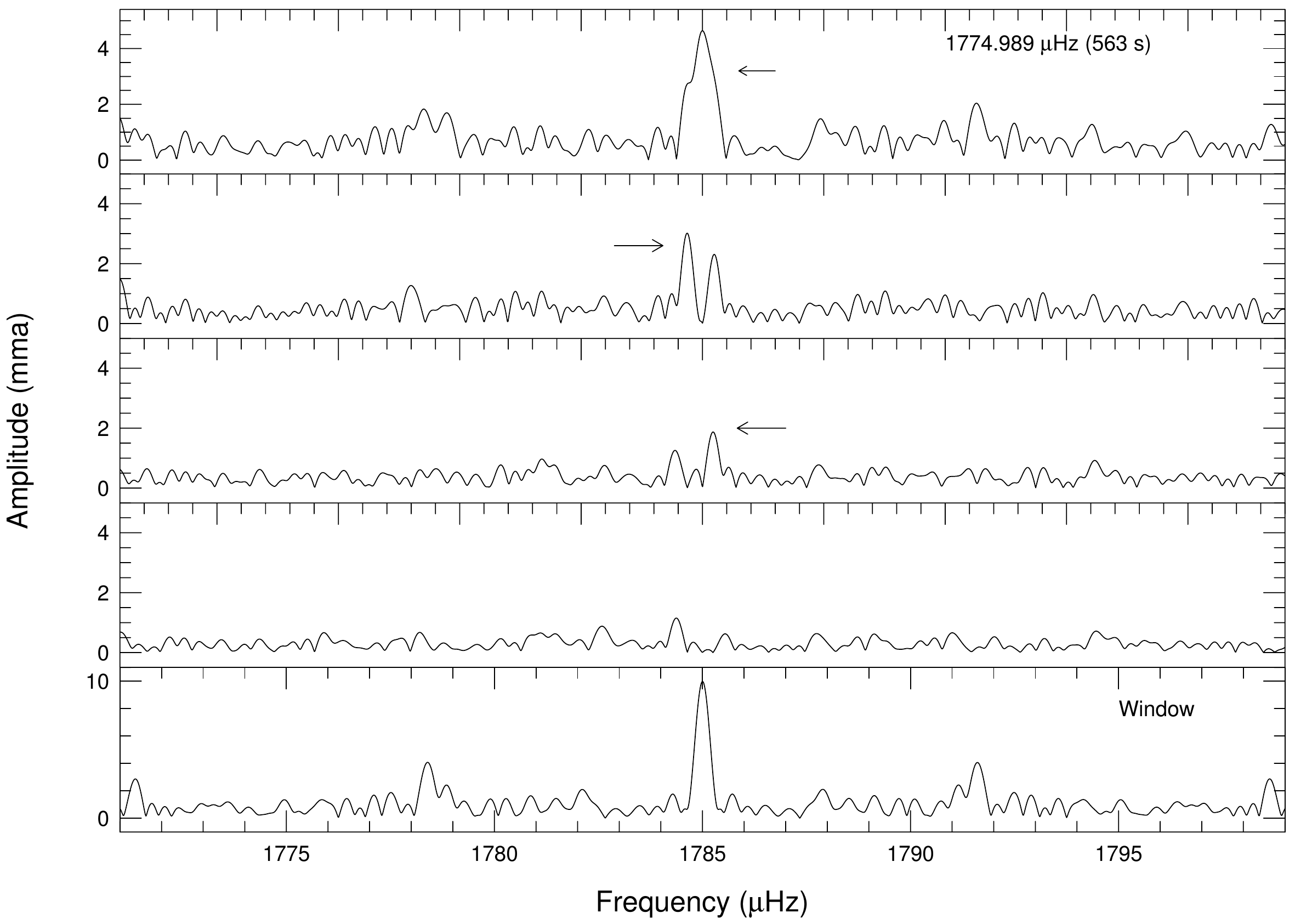}
}
\caption{The 1775 \muHz\ (563 s) mode in the 2008 FT.  This power is
  unresolved, and requires at least 4 closely space frequencies for
  prewhitening, with frequency differences very similar to the inverse
  of the run length. This is a clear signature of amplitude and/or
  frequency instability.\label{w1775}}
\end{figure}

Similar analysis of several other frequencies turned out to be
unexpectedly complex. The power at 1775 \muHz\ (labeled ``5'' in
Figure~\ref{xcov26dft}) is unresolved. Figure~\ref{w1775} establishes
that prewhitening this peak requires 4 closely spaced frequencies,
with frequency differences very similar to the inverse of the run
length. This is a clear signature of amplitude and/or frequency
instability. Figure~\ref{w1775spgm} shows the spectrogram of this
region using the same criteria as the spectrograms in
Figure~\ref{ec11allspgmcolor}. Unlike the slow amplitude increase
observed with the 1633 \muHz\ peak, the 1775 \muHz\ peak undergoes
remarkably sudden variations, on timescales of a few days. We also
find a a decrease in its frequency of 1.2 \muHz\ (5$\sigma$) over the
course of the run. A second region of power at 1860 \muHz\ displays
similar behavior. Both peaks are labeled in Table~\ref{tab:freq}.

Our final identifications result from a simultaneous nonlinear least
squares fit of 19 independent frequencies, amplitudes, and phases as
well as 68 combination frequencies. Combination frequencies are fixed with 
respect to their parents but their amplitudes and phases are allowed to 
vary. Table~\ref{tab:freq} lists 19 identified independent
frequencies, with consideration to those exhibiting amplitude and/or
frequency modulation. The list includes a doublet with a splitting of
9.9 \muHz\ associated with the 1633.907 \muHz\ peak (labeled 1 in
Figure~\ref{xcov26dft}), a triplet at 1887 \muHz\ with an average
splitting of 3.8 \muHz\ (labeled 2 in Figure~\ref{xcov26dft}) and a
second doublet with a splitting of 3.1 \muHz\ at 2504 \muHz\ (labeled
4 in Figure~\ref{xcov26dft}).  Table~\ref{tab:comb} presents the
largest amplitude combination frequencies (see
Section~\ref{sec:comfreq}).

\begin{figure}[b]
  \centering{
\includegraphics[width=1.0\columnwidth]{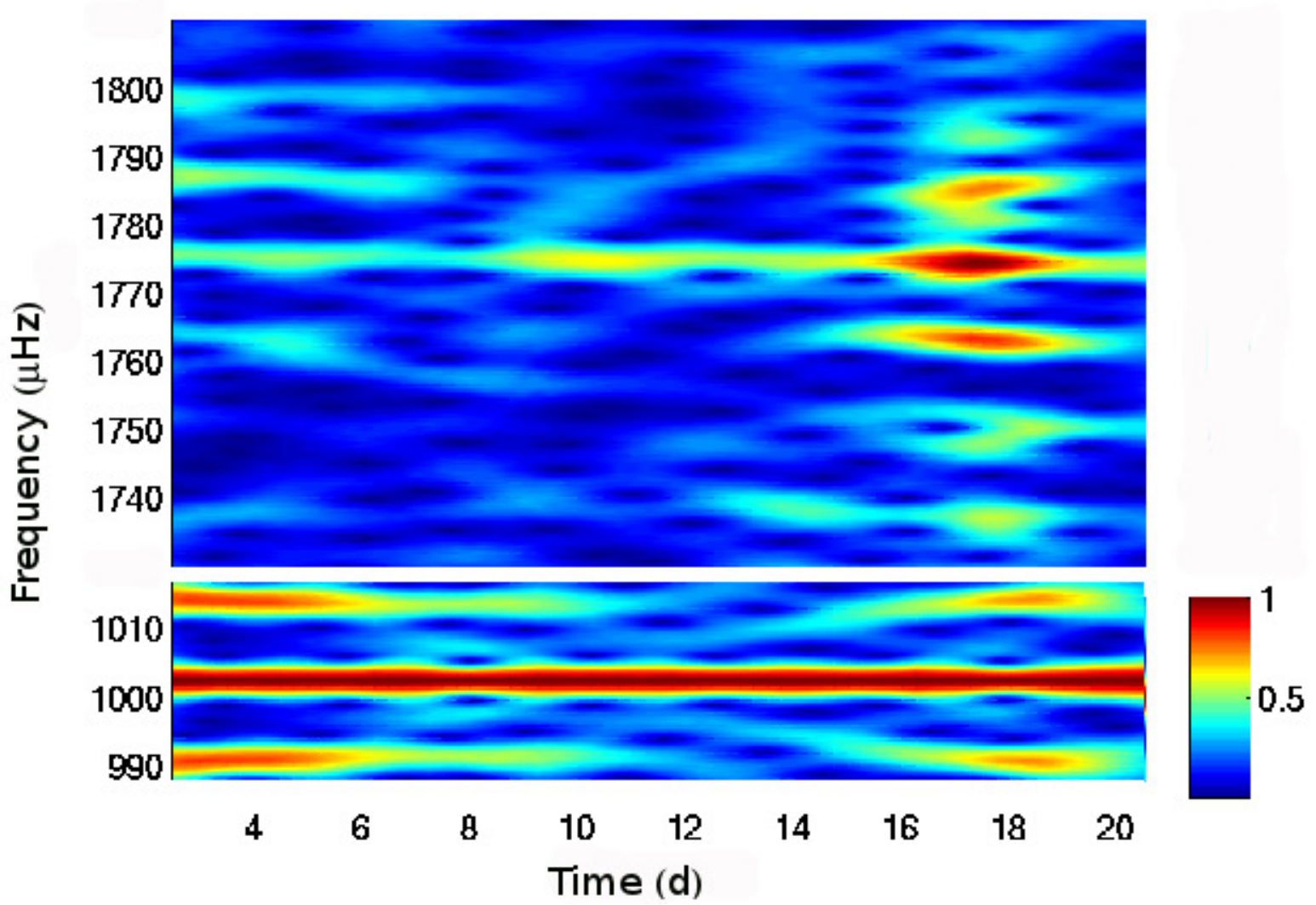}
}
\caption{Spectrogram of the 1775 \muHz\ region in the 2008 FT.  This
  region of power exhibits remarkable amplitude changes on a
  timescale of days. The bottom panel gives the time dependent window.  
  The spectrogram amplitudes (colors) have been normalized to a value of 1. 
  }
\end{figure}

\section{Mode identification}\label{sec:id}

\begin{deluxetable}{rrrrrrr} 
\tablecaption{Table of XCOV27 Independent Frequencies}
\tablewidth{0pt}
\tablehead{
\colhead{ID} & \colhead{Frequency} & \colhead{Period} & \colhead{Amplitude} & \colhead{S/N} &\colhead{Notes} \\
\colhead{} & \colhead{(\muHz)} & \colhead{(s)} &\colhead{(mma)} & \colhead{} &\colhead{} &\colhead{}
}
\startdata
7 & $935.3799\pm0.002$ & 1069.085 &$2.7\pm0.1$ & 9 \\
10 & $1021.139\pm0.002$ & 979.299 & $1.7\pm0.1$ & 8 \\
9 & $1104.252\pm0.001$ & 905.591 & $2.2\pm0.2$ & 10 \\
11 & $1155.925\pm0.002$ & 865.108 & $1.9\pm0.2$ & 8 \\
14 & $1241.403\pm0.002$ & 805.540 & $1.2\pm0.2$ & 5 \\
13  & $1418.369\pm0.002$ & 705.035 & $1.2\pm0.2$ & 5  \\
3a  & $1521.575\pm0.002$ & 657.214 & $2.2\pm0.2$ & 9 &\tablenotemark{**} \\
3 & $1548.146\pm0.001$ & 645.933 & $7.9\pm0.1$ & 32 \\
1a  & $1624.015\pm0.003$ & 615.758 & $3.1\pm0.2$ & 13 & \tablenotemark{**}\\
1 & $1633.907\pm0.001$ & 612.030 & $25.7\pm0.1$ & 104 \\
5  & $1774.989\pm0.100$ & 563.384 & $7.2\pm0.3$ & 9 & \tablenotemark{*}\\
8  & $1860.248\pm0.400$ & 537.563 & $6.4\pm0.4$ & 11 & \tablenotemark{*}\\
2a   & $1883.555\pm0.003$ & 530.911 & $1.5\pm0.2$ & 5 & \tablenotemark{**}\\
2  & $1887.404\pm0.001$ & 529.828 & $20.7\pm0.1$ & 79 \\
2b  & $1891.142\pm0.002$ & 528.781 & $3.8\pm0.2$ & 14  & \tablenotemark{**}\\
6 & $2304.745\pm0.001$ & 433.887 & $4.7\pm0.1$ & 23 &\tablenotemark{*}\\
4 & $2504.897\pm0.001$ & 399.218 & $12.7\pm0.1$ & 71  \\
4a  & $2508.060\pm0.002$ & 398.715 & $2.1\pm0.2$ & 12 & \tablenotemark{**}\\
12& $2856.155\pm0.002$ & 350.121 & $2.0\pm0.2$ & 9 
\enddata
\tablenotetext{*}{Frequencies displaying long timescale amplitude modulation}
\tablenotetext{**}{Frequency IDs with small letters indicate components of the same multiplet.}
\label{tab:freq}
\end{deluxetable}

Our ultimate goal is to use EC14012-1446's nonlinear light curve to
extract the star's convective parameters. The requirements for
convective light curve fitting include precise knowledge of the target
star's frequencies and amplitudes, as well as a good idea of the
($l,m$) values for these frequencies. The parameter space to be
explored to determine the best convective light curve fit is greatly
constrained by knowledge of ($l,m$) indices for the excited
pulsations. Montgomery's first application of this technique focused
on the DA pulsators G29-38 and GD154 \citep{Montgomery05a}. In the
data sets used, these objects were dominated by a single pulsation
mode, so the number of potential ($l,m$) identifications was small
enough that all possibilities could be explored. \citet{Montgomery10}
recently expanded this work to the multiperiodic pulsator GD358, a
well-studied object with detailed ($l,m$) identifications of its
pulsation modes (\citet{Provencal09}; \citet{Metcalfe00};
\citet{Winget94}).  While XCOV26 provided precise frequencies and
amplitudes for the modes in EC14012-1446 (see Table~\ref{tab:freq}),
we lack prior ($l,m$) identifications. Even limiting ourselves to
$l$=1 and 2, as these are the spherical degrees most often observed in
pulsating white dwarfs, this yields a total number of possible
combinations of order $(2l+1)^{19}$, a very large number indeed. In
this section, we will rely on analysis of the combination frequencies
and the support of asteroseismology to constrain ($l,m$)
identifications in EC14012-1446.

\subsection{Combination Frequencies}\label{sec:comfreq}

\begin{deluxetable}{rrrrr} 
\tablecaption{Table of Combination Frequencies}
\tablewidth{0pt}
\tablehead{
\colhead{Frequency} & \colhead{Period} & \colhead{Amplitude} & \colhead{S/N} & \colhead{Parents} \\
\colhead{} & \colhead{(\muHz)} & \colhead{(s)} &\colhead{(mma)} & \colhead{} 
}
\startdata
 $253.466\pm0.002$ & 3945.302 & $2.9\pm0.2$ & 9 & 2-1 \\
 $2569.286\pm0.002$ & 350.121 & $0.6\pm0.2$ & 4 & 1+7 \\
 $2738.153\pm0.002$ & 365.210 & $0.9\pm0.2$ & 4 & 1+9  \\
 $3096.255\pm0.002$ & 322.971 & $0.6\pm0.2$ & 3 & 2f3  \\
 $3182.045\pm0.002$ & 314.263 & $2.7\pm0.2$ & 12 & 1+3 \\
 $3267.794\pm0.002$ & 306.015 & $4.3\pm0.2$ & 13 & 2f1 \\
 \tablenotemark{*} $3408.939\pm0.002$ & 293.346 & $1.6\pm0.2$ & 5 & 6+5  \\
 \tablenotemark{*} $3409.912\pm0.002$ & 293.263 & $1.7\pm0.2$ & 4 & 1+5  \\
 $3435.538\pm0.002$ & 291.075 & $2.0\pm0.2$ & 6 & 2+3  \\
 \tablenotemark{*} $3494.064\pm0.002$ & 286.152 & $0.8\pm0.2$ & 4 & 1+8\\
 $3521.306\pm0.002$ & 283.986 & $7.0\pm0.2$ & 33 & 1+2 \\
 $3525.050\pm0.002$ & 283.684 & $0.9\pm0.2$ & 5 & 1+2b  \\
 $3774.810\pm0.001$ & 264.914 & $2.1\pm0.2$ & 12 & 2f2  \\
 $3938.646\pm0.002$ & 253.894 & $0.9\pm0.2$ & 5 & 1+6  \\
 $4026.445\pm0.002$ & 248.358 & $1.4\pm0.2$ & 7 & 4+3a \\
 $4053.043\pm0.002$ & 246.728 & $0.7\pm0.2$ & 4 & 3+4 \\
 $4138.800\pm0.002$ & 241.616 & $2.0\pm0.2$ & 8 & 1+4 \\
 $4192.150\pm0.002$ & 238.541 & $0.7\pm0.2$ & 4 & 2+6 \\
 \tablenotemark{*} $4279.907\pm0.002$ & 233.650 & $2.6\pm0.2$ & 10 & 4+5 \\
 \tablenotemark{*} $4365.040\pm0.002$ & 234.464 & $0.9\pm0.2$ & 4 & 4+8  \\
 \tablenotemark{*} $4365.608\pm0.002$ & 229.063 & $1.3\pm0.2$ & 5 & 4+8  \\
 $4392.348\pm0.002$ & 227.669 & $1.5\pm0.2$ & 7 & 2+4  \\
 $4396.055\pm0.002$ & 227.477 & $1.8\pm0.2$ & 8 & 2b+4  \\
 $4743.561\pm0.002$ & 210.812 & $0.9\pm0.2$ & 5 & 2+12 \\
 $4809.633\pm0.002$ & 207.916 & $2.2\pm0.2$ & 12 & 4+6 \\
 $5009.696\pm0.002$ & 199.613 & $0.7\pm0.2$ & 4 & 2f4  \\ 
 $5155.187\pm0.002$ & 193.979 & $2.2\pm0.2$ & 8 & 1+1+2  \\
 $5361.052\pm0.002$ & 186.531 & $1.2\pm0.2$ & 6 & 3+10  \\
 $5408.698\pm0.002$ & 184.887 & $1.3\pm0.2$ & 6 & 3+3+1 \\
 $5913.800\pm0.002$ & 169.096 & $1.4\pm0.2$ & 7 & 1+4+5
\enddata
\label{tab:comb}
\tablenotetext{*}{Combination including parent with amplitude modulation}
\end{deluxetable}

Combination frequencies are typically observed in the FTs of moderate
to large amplitude pulsators \citep[e.g.,][]{Provencal09, Dolez06,
  Handler02}; they are identified by their relationships, which must
be exact within measurement error. Combination frequencies can be
integer multiples of a single parent (harmonics), or sums (or
differences) of any two modes. These frequencies are not independent,
but result from nonlinear effects most likely associated with the
surface convection zone \citep{Brickhill92a, Brassard95a, Wu01,
  Ising01}. \citet{Wu01} shows that the observed amplitudes of
combination frequencies depend on geometric factors such as the
($l,m$) indices of the parent(s) and the inclination of the pulsation
axis to the line of sight. EC14012-1446's FT contains a rich
distribution of combinations that involve 13 of the 19 independent
frequencies listed in Table~\ref{tab:freq}. Our goal is to exploit the
geometric sensitivities to provide ($l,m$) constraints for these
frequencies.

\citet{Wu01} and \citet{Yeates05} lay the foundations for our analysis
by describing analytical expressions for the predicted amplitudes and
phases of combination frequencies. These quantities depend on the
inclination angle $\theta$ of the pulsation axis to the line of sight,
the ($l,m$) indices of the parent mode(s), the amplitudes of the
parent mode(s), and parameters describing the convection zone. To
minimize the dependence on the convective parameters and focus on the
geometric factors ($l,m$) and $\theta$, our analysis follows
\citet{Yeates05} and considers only combination frequencies that are
the sum of two parent frequencies. We combine a genetic algorithm
\citep{Charbonneau95} with \citet{Wu01}'s formulae, and incorporate an
improved treatment of limb darkening taken directly from the models of
Koester \citep{Montgomery10}. For a single run of the code, the best
simultaneous fit to the observed amplitudes of the parent and
combination frequencies utilizes multiple generations and minimizes
the root-mean-squared residuals, Res$_{\rm rms}$, between the predicted and
observed amplitudes. In practice, we run the code 1000 times, and
select the solutions having Res$_{\rm rms}$ below a limiting value.  This
process produces a sample of best-fit solutions whose distribution
provides information on the range of values allowed for these
parameters.

To test that this approach recovers known input, we used the nonlinear
light curve fitting code of \citet{Montgomery10} to generate a
synthetic light curve based on EC14012-1446's parent frequencies. The
nonlinear light curve fitting code is discussed in more detail in
Section~\ref{sec:nonlinear}. For our purposes in this test, the
importance of using synthetic light curves generated by this code
rather than a simple simulation employing multiple sine functions is
that the synthetic light curve will include combination frequencies
produced via nonlinear effects due to convection
\citep{Montgomery10}. Our purpose is to recover the input frequency
identifications using these combination frequencies. We assigned
reasonable ($l,m$) values to the input frequencies, chose values for
$\rm{\theta}$, the time-averaged convective response time $\tau_0$, 
and \A, where \A\ describes the response of
the stellar material to the pulsations and includes a bolometric
correction factor \citep{Wu01}, and added noise. Using the method
outlined in the previous paragraph, we successfully recovered ($l,m$)
for all large amplitude input parent frequencies, while experiencing
some disagreement with low amplitude parents. This is because low
amplitude modes have even lower amplitude combination frequencies,
which are difficult to detect and so are not as numerous.
Since our method treats both the high amplitude parent modes and the
low amplitude combination frequencies equally in the fits, this has
the effect of de-emphasizing the importance of low amplitude
combination frequencies.

\begin{figure}
  \centering{
\includegraphics[width=1.00\columnwidth]{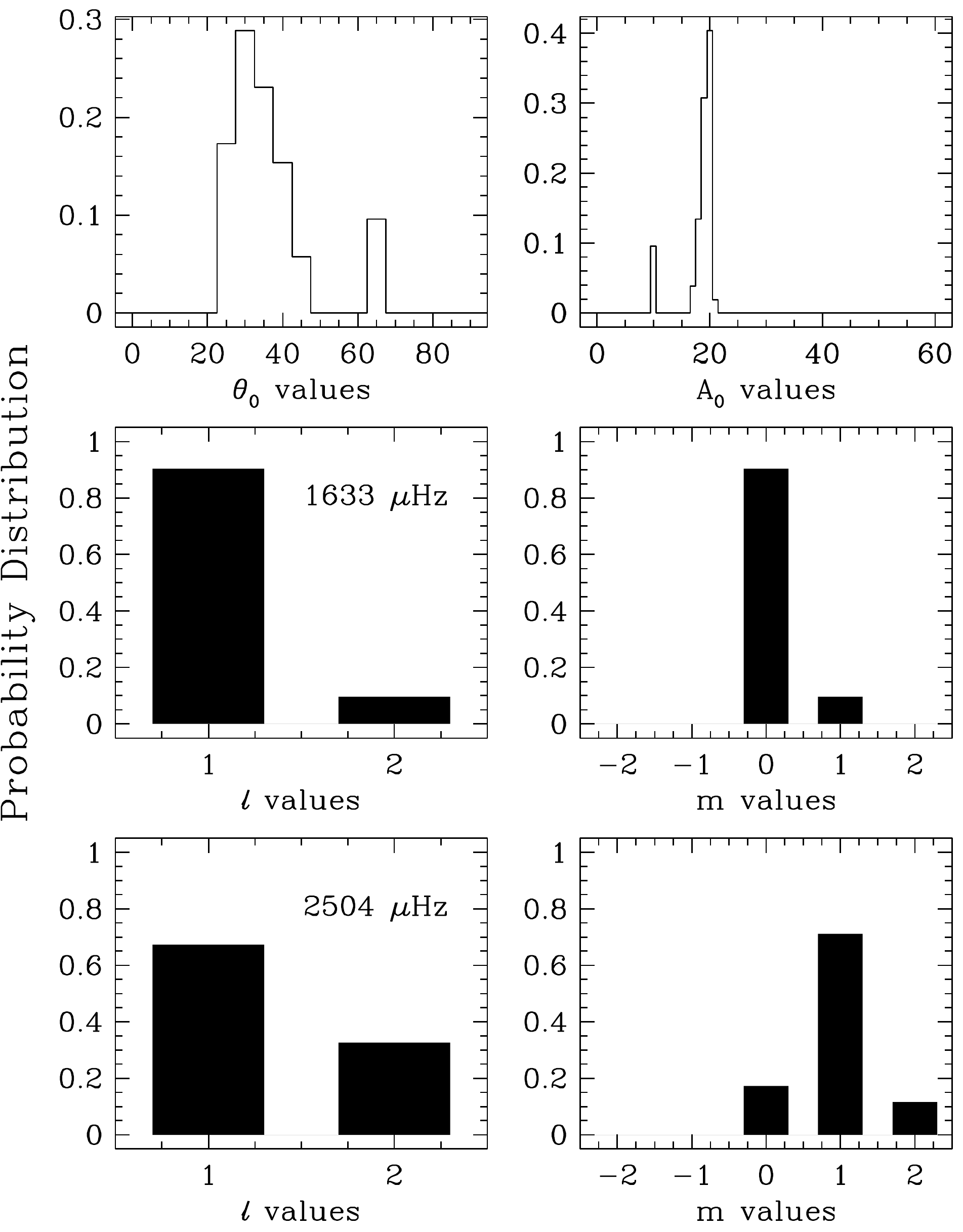}
}
\caption{Probability distribution (from 0--1) of $l$, $m$, ${\theta}$,
  and \A\ for the 1633.907 \muHz\ (612 s) and the 2504.897 \muHz\ (399
  s) variations. The solutions are from individual runs with Res$_{\rm
    rms} <0.38$~mmi.  The top panel gives the probability
  distributions of $\theta$ and \A\ found for all frequencies, with
  preferred values of $\theta \sim \rm{32^\circ}$ and \A$ \sim 20$,
  respectively. The middle and lower panels give the $l$ and $m$
  identification distributions for the 1633.907 \muHz\ (612 s) and
  2504.897 \muHz\ (399 s) mode.  The amplitudes of EC14012-1446's
  observed combination frequencies argue that 1633.907 \muHz\ is
  $l$=1, $m$=0, while 2504.897 \muHz\ is $l=1$, $m=1$.}
  \label{hist}
\end{figure}


\begin{deluxetable*}{rrrrr}
\tablecaption{Frequency ($l,m$) Identifications}
\tablewidth{0pt}
\tablehead{
\colhead{Frequency} & \colhead{Period} & \colhead{Combination} & \colhead{Asteroseismology} &\colhead{Final Nonlinear Fit}  \\
\colhead{\muHz} & \colhead{(s)} & \colhead{($l,m$)}
&\colhead{($l,m$)} & \colhead{($l,m$)}}
\startdata
$935.3799\pm0.002$ & 1069.085 &  &(1,any) & (2,-1)\\
$1021.139\pm0.002$ & 979.299 & &(1,any) & (2,2)\\
$1104.252\pm0.001$ & 905.591 &  &(1,any) & (2,1)\\
$1155.925\pm0.002$ & 865.108 &  &(1,any) & (2,2)\\
$1241.403\pm0.002$ & 805.540 &  & (1,any)& (2,2) \\
$1418.369\pm0.002$ & 705.035 &  &(2,any)  & (2,1)\\
$1521.575\pm0.002$ & 657.214 &  & (1,1) & 1,1)\\
$1548.146\pm0.001$ & 645.933 & (1,0) & (1,any) & (1,0)\\
$1624.015\pm0.003$ & 615.758 & (1,1) & (1 1) & (1,1)\\
$1633.907\pm0.001$ & 612.030 & (1,0) & (1,0/1) & (1,1)\\
$1774.989\pm0.100$ & 563.384 & (1,-1) & (1,any) & (1,1)\\
$1860.248\pm0.400$ & 537.563 & &(2,any)  & (2,0)\\
$1883.555\pm0.003$ & 530.911 &  & (1,1)& (1,1)\\
$1887.404\pm0.001$ & 529.828 & (1,0) & (1,0)  & (1,0)\\
$1891.142\pm0.002$ & 528.781 & (1,-1) & (1,-1) & (1,-1)\\
$2304.745\pm0.001$ & 433.887 & (1,-1) & (2,any)  & (1,0)\\
$2504.897\pm0.001$ & 399.218 & (1,1) & (1,1) & (1,0) \\
$2508.060\pm0.002$ & 398.715 & (1/2,1/0) & (1,0) & (1,-1)\\
$2856.155\pm0.002$ & 350.121 & (2,0/1)& (2,any) & (2,0)
\enddata
\label{tab:freqid}
\end{deluxetable*}

Figure~\ref{hist} shows the resulting probability distribution of $l$,
$m$, ${\theta}$, and \A\ values for fits having Res$_{\rm rms}
<0.38$~mmi, for EC14012-1446's 1633.907 \muHz\ (612 s) and 2504.897
\muHz\ (399 s) variations. For 1633.907 \muHz\ (612 s), analysis of
combination amplitudes strongly argues that the variations are best
represented as spherical degree $l$=1 and azimuthal index $m$=0, with
an inclination angle of $30\pm10^\circ$, and a value of \A$
=20\pm2$. We find nearly identical results for 1887.404 \muHz\ (529 s)
and 1548.146 \muHz (646 s). For 2504.897 \muHz (399 s), we find the same distributions
for $\rm{\theta}$ and \A, but in this case $(l,m)$=(1,1) is strongly
preferred. As we found previously, the statistical significance of
$(l,m)$ identifications determined by combination amplitudes is
amplitude dependent and we do not find unambiguous identifications for
the lower amplitude frequencies. The complete list of $(l,m$)
identifications derived from the combination amplitudes are given in
Table~\ref{tab:freqid}.

Our analysis of EC14012-1446's combination amplitudes indicates that 9
of the dominant independent frequencies are consistent with spherical
degree $l$=1; in Table~\ref{tab:freqid} we list the most likely ($l$,
$m$) identifications for these modes. We would like to constrain the
identifications of the remaining independent frequencies to improve
our chances of success with nonlinear convective light curve
fitting. Using the combination analysis as a basis, we now turn to
asteroseismology to provide constraints on the identifications of the
remaining frequencies.

\subsection{Asteroseismology}\label{sec:astero}

\subsubsection{Period Spacing}
As mentioned in the Introduction, we look for two clues to indicate
($l,m$) identifications of pulsations in white dwarfs. The first 
is the expectation that g-mode pulsations of a given $l$ corresponding to 
successive radial overtones $k$ will be approximately equally spaced in 
period, provided $k$ is large enough \citep{Unno89}. 
For the simple example of a homogeneous star, we find
\begin{equation}
{P_{kl}=k\,\,\Delta \Pi \,\,[l(l+1)]^{-1/2} \, + \, C}
\end{equation}
where $\Delta\Pi$ is a uniform period spacing, ${P_{kl}}$ is the
period for a given mode ($k,l$), and $C$ is a constant
\citep{Tassoul90}.  The mean period spacing $\Delta\Pi$ for a series
of modes of a given $l$ and consecutive $k$ is an important
asteroseismic measure of stellar mass and effective temperature that
is mostly independent of internal composition \citep{Unno89}.

\begin{figure}[b]
  \vspace*{0.5em}
  \centering{
\includegraphics[width=1.0\columnwidth]{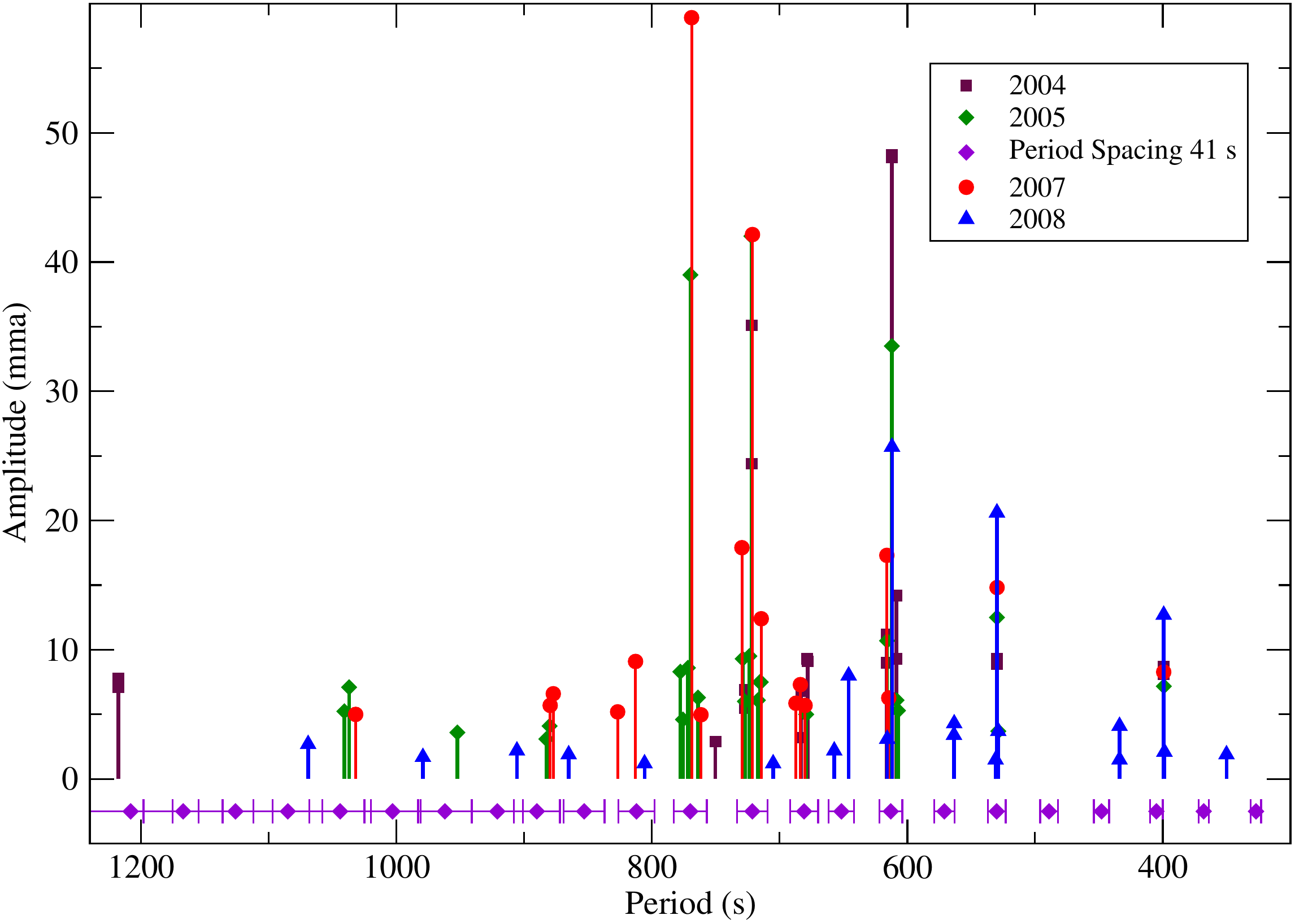}
}
\vspace*{-0.8em}
\caption{Schematic representation of EC14012-1446's independent
  frequencies for data from 2004--2008.  The bottom points indicate
  the expected locations of consecutive $l=1$ modes with a mean period
  spacing of 41 s. The associated bars illustrate the ``grouping
  width'' of $\pm 24$ \muHz. Frequencies within this range should
  belong to the given $k$ multiplet. Note that the $x$-axis is given
  in period, not frequency. 
  \label{periodplot}}
\end{figure}

\begin{figure}[b]
  \vspace*{0.5em}
  \centering{
\includegraphics[angle=0,width=1.0\columnwidth]{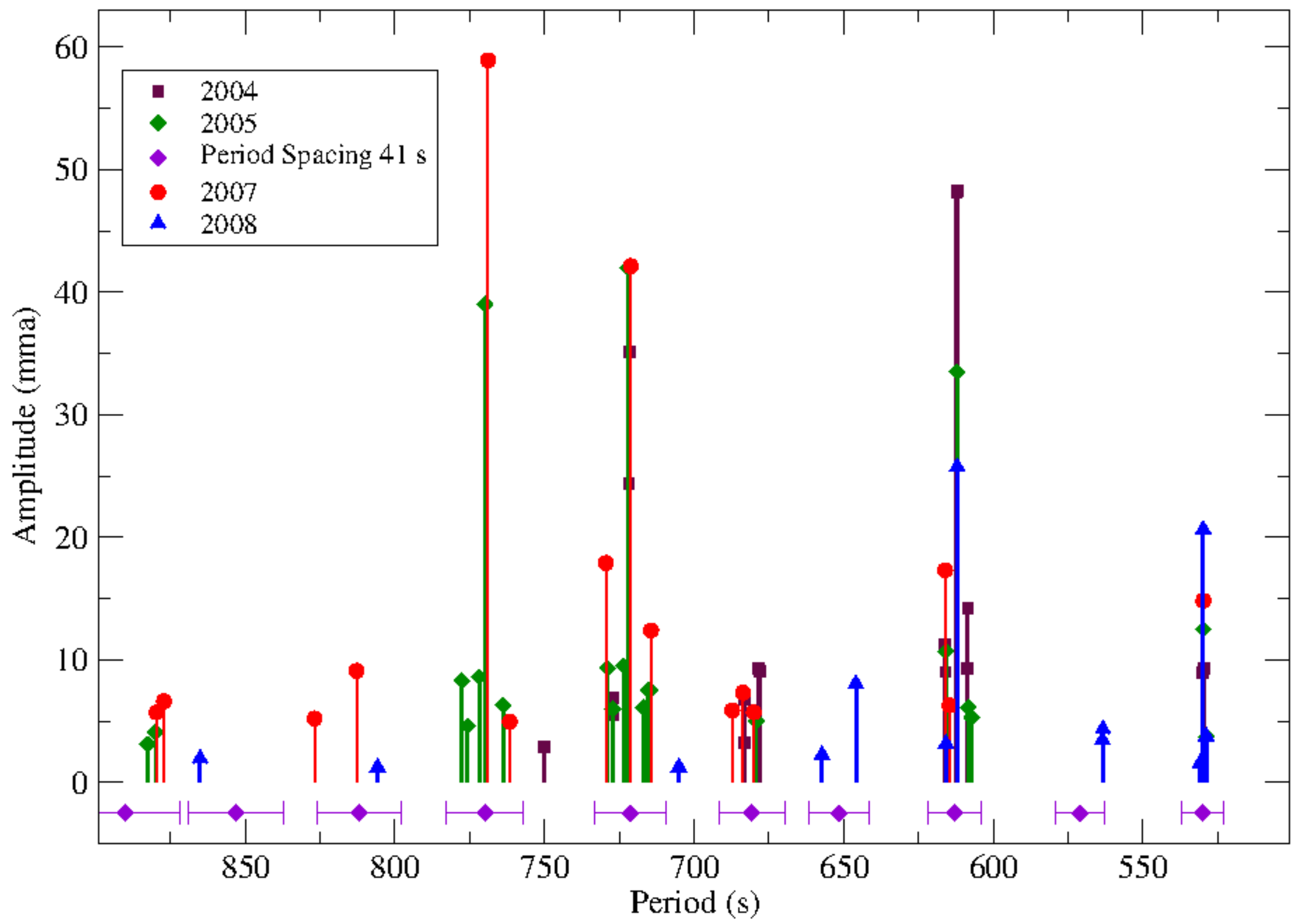}
}
\vspace*{-0.8em}
\caption{A zoom-in of Figure~\ref{periodplot} to periods in the range
  800--500 s.  The bottom points give
  the expected locations of consecutive $l$=1 modes with a mean period
  spacing of 41 s. The period width denoted by the bars represent the
  24 \muHz\ ``grouping width'' of the modes. Frequencies found within
  this range should belong to the given $k$ multiplet. Note that the
  $x$-axis is given in period, not frequency. 
  \label{periodplot2}}
\end{figure}

A white dwarf has a host of available pulsation frequencies. For
reasons that are not understood, in most cases only a subset are
observed at any given time \citep{Corsico02}. This is true for
EC14012-1446. The 19 independent frequencies detected during XCOV28
show no obvious evidence of equal period spacing so we are lacking
consecutive radial overtones for any $l$ value. A good strategy to
identify the complete set of available modes is to combine results
from multiple seasons of observations. \citet{Handler08} present an
analysis of EC14012-1446 observations spanning 2004--2007, during
which the star exhibited different subsets of excited
modes. Figure~\ref{periodplot} shows a schematic representation of
excited modes for the combined observations spanning
2004--2008. Figure~\ref{periodplot2} focuses on the obvious groupings
between 800 and 500 s (1250 and 2000 \muHz).  Since we are searching
for equal period spacings, we present these figures using period as
the $x$-axis. The large groupings at 768 s (1302 \muHz), 721 s (1387
\muHz), 682 s (1466 \muHz), and 612 s (1633 \muHz) have a decreasing
observed width with shorter period. This decrease translates into an
equal width in frequency space of $\approx 24$~\muHz.

\begin{deluxetable}{rrcc} 
\tablecolumns{4}
\tablewidth{17pc}
\tablecaption{Table of Average Frequencies (2004--2008)}
\tablehead{
\colhead{Period} & \colhead{Frequency} & \colhead{No. of peaks} & \colhead {radial overtone} \\
\colhead{(s)} & \colhead{(\muHz)} &\colhead{} 
}
\startdata
2856.164 & 350.120 & 1\\
2738.076& 365.220 & 1 & ${k_{\rm o}-6}$ \\
2505.405 & 399.137 & 6 & ${k_{\rm o}-5}$ \\
2304.854 & 433.865 & 2 \\
1887.505 & 529.800 & 10\phantom{1}& ${k_{\rm o}-2}$ \\
1860.431 & 537.510 & 2 \\
1774.954 & 563.395 & 2 & ${k_{\rm o}-1}$\\
1633.597 & 612.146 & 16\phantom{1} & ${k_{\rm o}}$\\
1548.155 & 645.930 & 2 & ${k_{\rm o}+1}$\\
1463.668 & 683.215 & 8 & ${k_{\rm o}+2}$\\
1418.353 & 705.040 & 1 \\
1384.903 & 722.072 & 14\phantom{1} &${k_{\rm o}+3}$\\
1332.851 & 750.249 & 1 \\
1298.863 & 769.904 & 6 & ${k_{\rm o}+4}$\\
1219.836 & 819.782 & 2 & ${k_{\rm o}+5}$\\
1163.163 & 859.710 & 2 & ${k_{\rm o}+6}$\\
1136.538 & 879.865 & 4 & ${k_{\rm o}+7}$\\
1104.240 & 905.600 & 1 \\
1035.443 & 965.770 & 2 & ${k_{\rm o}+9}$\\
964.748 & 1036.540 & 3 & ${k_{\rm o}+11}$\\
935.375 & 1069.090 & 1 \\
821.254 & 1217.650 & 1
\enddata
\label{tab:avg}
\tablecomments{This table gives the simple average frequencies for the
  groupings in Figure~\ref{periodplot} and
  Figure~\ref{periodplot2}. The column ``No. of peaks'' gives the
  number of frequencies contained within each group. The radial
  overtone $k$ is impossible to determine observationally, so we give
  values relative to 1633~\muHz, defined to have $k_{\rm o}$. The
  relative $k$ values denote a series of nearly consecutive modes of
  spherical index $l$=1 with a period spacing of $\approx 41$~s.}

\end{deluxetable}

\begin{figure}[t]
  \centering{
\includegraphics[width=1.0\columnwidth]{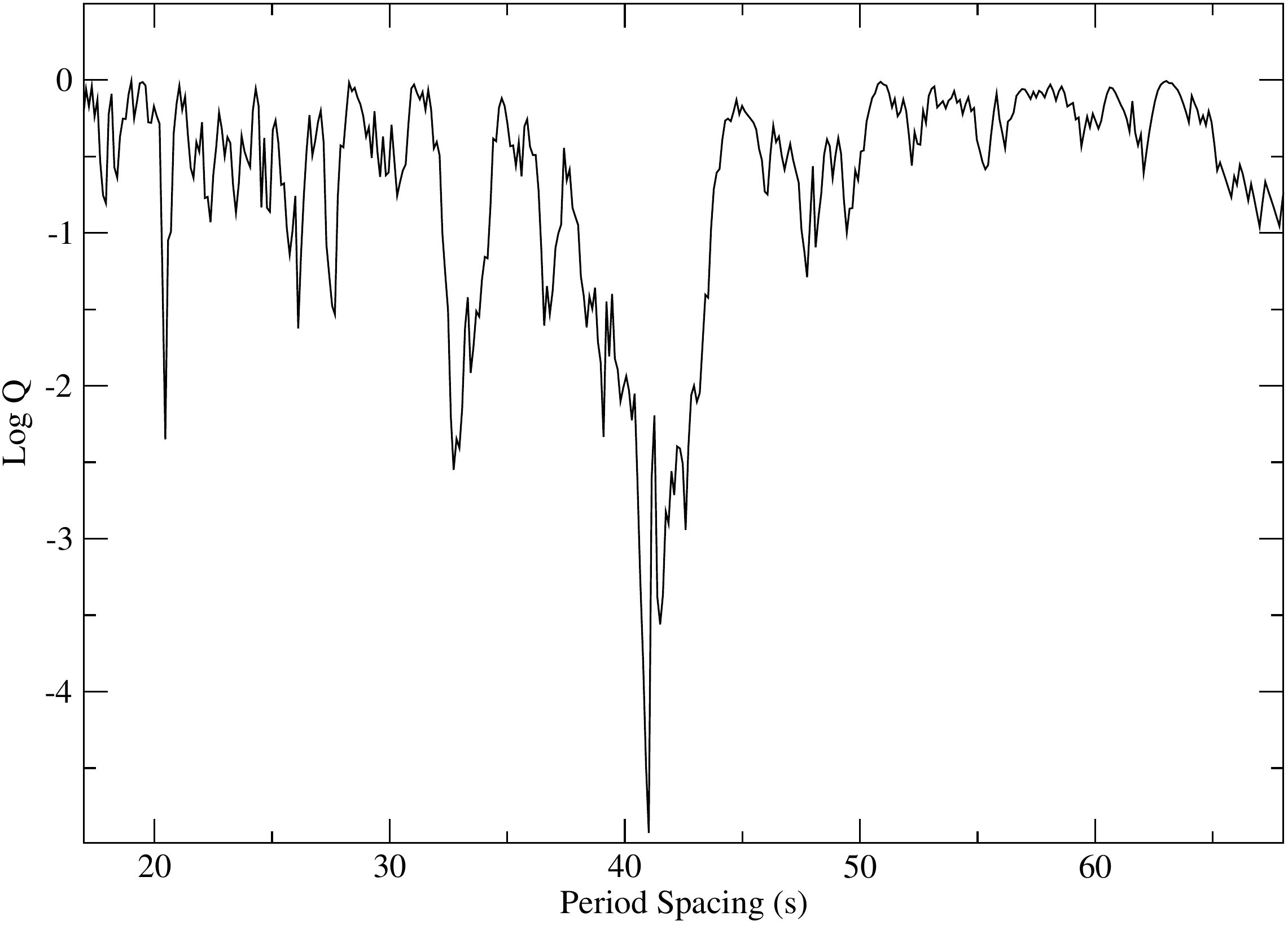}
}
\caption{Kolmogorov-Smirnov (KS) test applied to the average periods
  in Table~\ref{tab:avg}. A period spacing is defined as significant
  with a confidence level of $(1-Q)\times 100\%$. The results reveal
  an average period spacing of $\approx 41$~s with a confidence level
  of 99.99\%. \label{kstest}}
\vspace*{0.5em}
\end{figure}

We calculated a simple average period for each group, which is given
in Table~\ref{tab:avg}. A statistical test for the presence of uniform
period spacing is provided by the Kolmogorov-Smirnov (KS) test
\citep{Winget91a}. The KS test calculates the probability that an
input list is randomly distributed. Any statistically nonrandom period
spacing will therefore appear as a minimum value in the output. In our
case, we use the test to determine the probability that the list of
average periods in Table~\ref{tab:avg} are from a uniform distribution
for a given period spacing $\Delta\Pi$
\citep{Kawaler88}. Figure~\ref{kstest} shows the results, finding a
period spacing of $\approx$ 41 s, consistent with the expectations for
$l$=1 \citep{Kim2011}. We find no significant period spacings at
$\approx$23 s as predicted for $l$=2 modes.

Using 1633.907 \muHz\ as our reference point, we assigned relative
radial overtone values to the frequencies listed in Table 5. We plot
the expected locations of consecutive radial overtones for $l = 1$ in
Figures~\ref{periodplot} and \ref{periodplot2}.  These points in the
bottom of both figures are given with bars representing the 24 \muHz\
``grouping'' width. For periods above 900 s, the grouping widths for
consecutive radial overtones overlap, illustrating the difficulty of
assigning relative $k$ values to modes with periods longer than this
value. Nonetheless, we find that most of EC14012-1446's pulsation
frequencies can be identified as $l$=1.

\begin{figure}[b]
\vspace*{0.7em}
  \centering{
\includegraphics[width=1.0\columnwidth]{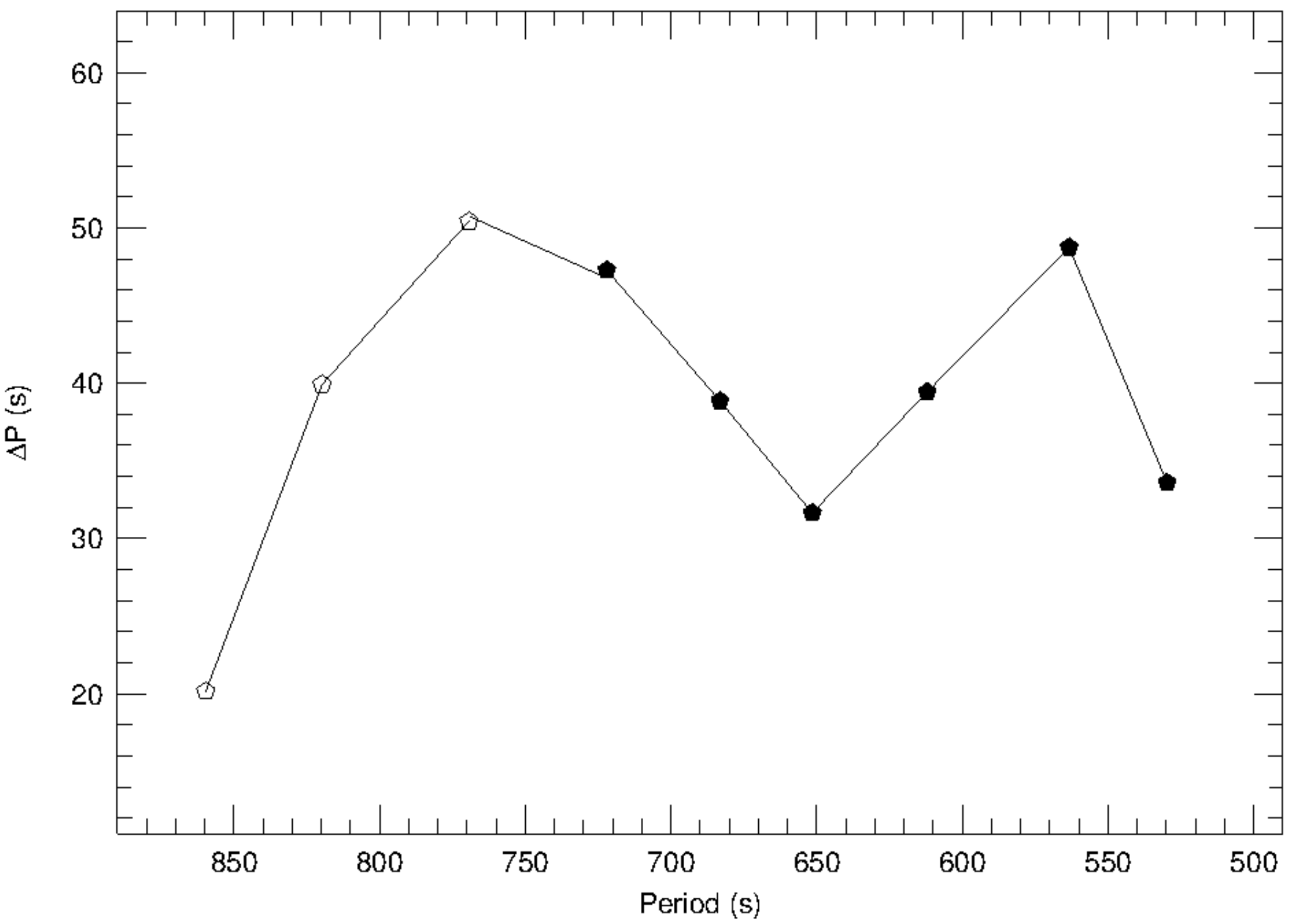}
}
\caption{$\Delta P$ vs $P$ for EC14012-1446, using the average periods
  from Table~\ref{tab:avg}. We use forward differencing (${\Delta
    P=P_{k}-P_{k+1}}$), where the radial overtone $k$ increases to the
  left. The solid points represent those frequencies with $l$=1
  identifications.  The open points represent those where our
  identification is uncertain.\label{deltap}}
\end{figure}

Unlike our simple example of a homogeneous star, white dwarfs are
compositionally stratified, so the individual period spacings will not
be uniform.  We can retrieve detailed information about interior
structure from the distribution of excited pulsation frequencies.  A
long standing problem with the asteroseismology of DA pulsating white
dwarfs is the lack of objects with rich pulsation spectra
\citep{Kim11}. In this respect, EC14012-1446 immediately reveals its
potential. Using our calculated average periods from
Table~\ref{tab:avg}, we show how individual period spacings $\Delta P$
differ as a function of period (and relative $k$ value) in
Figure~\ref{deltap}. We use ``forward differencing'', where ${\Delta P}$ 
is defined as ${\Delta P = P_{k}-P_{k+1}}$.  The filled points
represent the periods between $\approx 900$--500~s that we are certain
are consecutive radial overtones of $l$=1. The open circles represent
those periods above $\approx 900$~s with ambiguous $k$
identifications. The roughly cyclic behavior in Figure~\ref{deltap} is
a sign of mode trapping. Mode trapping occurs in compositionally
stratified stars when there is a resonance between a pulsation
frequency and a surface layer.  In theoretical models, a resonance
occurs when a radial node coincides with a transition layer. In a DA
white dwarf, transition zones occur at the boundary of the hydrogen
and helium layers, the helium layer and the carbon/oxygen core, and at
points in a possibly chemically stratified core \citep{Montgomery05c}.
The trapping cycle (the number of frequencies from minimum to minimum)
is most sensitive to the location of the trapping layer, e.g., the
base of the surface hydrogen layer. In addition, the trapping
amplitude (the depth of the minima) is sensitive to the density
gradient in the composition transition zone. In general, a larger
gradient produces a larger trapping amplitude. Work is underway to
determine EC14012-1446's detailed mass and internal structure
\citep{Kim2011}.

\subsubsection{Multiplet Structure}

A second clue to constrain ($l,m$) identifications of pulsations in
white dwarfs is the presence of rotationally split multiplets. The
multiplet components have the same $(k,l)$, and are further described
by the azimuthal index $m$, which takes integer values between $-l$
and $l$.  To first order, the frequency difference relative to the
$m=0$ component of the multiplet is given by
$\delta\nu_{klm}=-m\,\Omega\,(1-C_{kl})$, where $\Omega$ is the
rotation period and $C_{kl}$ is a coefficient that depends on the
pulsation eigenfunctions evaluated in the nonrotating case. In the
high-$k$ asymptotic limit for $g$-modes, $C_{kl} \sim 1/\ell
(\ell+1)$, although models predict it to vary by $\approx 10$ percent
over the range of observed periods in EC14012.  We adopt the
convention that a positive value of $m$ represents a prograde mode;
retrograde modes are denoted by negative $m$ values \citep{Winget94}.
An additional important diagnostic is given by the ratio of splittings
for $l$=1 and $l$=2, where we expect
$\rm{\delta\nu_{\ell=1}/\delta\nu_{\ell=2}}\simeq 0.6$ for g-modes
\citep{Winget91a}.

For EC14012-1446 during XCOV26, we find one triplet (1887 \muHz) and
two doublets (1633 and 2504 \muHz) among our list of 19 frequencies
(Table~\ref{tab:freq}).  We begin by assuming that the 1887 \muHz\
triplet is a rotationally split multiplet. Our combination analysis
argues that the central component (1887.404 \muHz, 529.828 s) is
($1,0$) and the 1891.142 \muHz\ (528.781 s) component is ($1,-1)$. The
1883.555 \muHz\ (530.911 s) component is low amplitude and is not
identified as a parent of any detected combination frequencies. We
cannot constrain its ($l,m$) identification via that method.  Given
($l,m$) identifications for the first two modes, the 1883.555 \muHz\
component should be ($1,1$). The average multiplet splitting of 3.8
\muHz\ implies a rotation period as sampled by these modes of
$1.53\pm0.01$~days.  This is a perfectly reasonable result, as
spectroscopic studies of white dwarfs reveal upper limits of $v\ \sin\
i\ = 10$~km~s$^{-2}$, consistent with rotation periods of order days
or longer \citep{Berger05}.

In the limit of uniform slow rotation, we expect additional $l=1$
multiplets to exhibit similar splittings and $l$=2 modes to have
splittings near 6.3 \muHz\ \citep{Winget91a}. Recalling that every $m$
component is not necessarily excited, we turn to the doublet at 1633
\muHz\ (612 s). Period spacing argues that this doublet is $l$=1, and
our combination analysis indicates that the 1633.907 \muHz\ (612.03 s)
mode is ($1,0$). The 1624.015 \muHz\ (615.758 s) component is not
found as a parent of any detected combination frequencies. If 1633.907
\muHz\ is indeed the central component of an $l$=1 triplet, then
1624.015 \muHz\ is $m=1$ and we do not detect the $m=-1$
component. However, the splitting of 9.89 \muHz\ differs significantly
from the expected value of $\approx 3.8$~\muHz\ found for the 1887
\muHz\ triplet. A second possibility is that the combination analysis
is incorrect and we are missing the multiplet's central component. In
this case, the 1633.907 \muHz\ mode is (1,-1) and the 1624.015 \muHz\
is ($1,1$), resulting in an average splitting of 4.95 \muHz\
($9.89/2$).  Moving to the 2504 \muHz\ (399 s) doublet, period spacing
again argues that this multiplet is $l$=1. Our combination analysis
indicates that the 2504.987 \muHz\ (399.204 s) component is
($1,1$). Again, the 2508.060 \muHz\ (398.715 s) component is not
identified as a parent of any detected combination frequency. The
frequency splitting between the two components is 3.16 \muHz\, and by
comparison with the splitting of the 1887 \muHz\ triplet, we can argue
that these modes have consecutive $m$ values, identifying the 2508.060
\muHz\ component as ($1,0$).

Multiplet splittings may also be used to eliminate the possibility
that EC14012-1446's frequencies represent a mixture of $l$=1 and $l$=2
modes. Assuming the 1887 \muHz\ triplet is $l$=1, we do not find
evidence for $l$=2 splittings of $\rm{\delta\nu_{l=2}=(3.8/0.6)=}$6.3
\muHz\ as predicted in the limit of slow, uniform rotation. Although
both the period spacing and the combination analysis argues against
it, to play devil's advocate we consider the possibility that the 1887
\muHz\ triplet is actually an $l=2$ quintuplet, since all $m$
components are not necessarily excited to observable levels. In this
case, then the expected $l=1$ multiplet splitting is
$\delta\nu_{l=1}=3.8 \times 0.6 = 2.3~\muHz$.  We find no examples of
multiplet splittings near this value. Finally, we must consider that
$\delta\nu_{3.16}/\delta\nu_{4.95}=0.64$.  Based on multiplet
structure alone, we could argue that the 1633.907 \muHz\ frequency is
$l=2$, and 2504.897 \muHz\ is $l=1$, but this is not supported by
either the period spacings or the combination analysis and leaves no
clear explanation for the 1887 \muHz\ triplet.

Clearly, the multiplets found in EC14012-1446's XCOV26 data set are
not well explained based on the simple model of rotational splitting.
Multiplet structure should be determined by the star's rotation rate
and structure.  We expect this to remain unchanged over time.  The
classical example is PG1159-035, which exhibits triplets and
quintuplets corresponding to $l=1$ and $l=2$
\citep{Winget91a}. However, complex multiplets are not unusual for
white dwarf pulsators.  In some instances, the assumption of rigid
rotation is clearly violated \citep{Corsico2011}.  In cooler pulsators
with moderate amplitudes and well developed convection zones, observed
multiplet structure can exhibit complicated behavior.  For example,
\citet{Provencal09} show changes in the DBV GD358's multiplet
structure that cannot be explained with simple rotational splitting.
For this same star, \citet{Winget94} show a dependence of multiplet
splittings that is not explained by expected variations of
$C_{kl}$. Processes that may play a role in multiplet structure of
cooler pulsators include changing weak magnetic fields similar to the
solar cycle, oblique pulsation, and differential rotation.  Our XCOV26
data on EC14012-1446 provides us with a single snapshot of this star's
multiplet structure. We need more observations to understand their
behavior.

\begin{figure}[t]
 \centering{
\includegraphics[width=1.0\columnwidth]{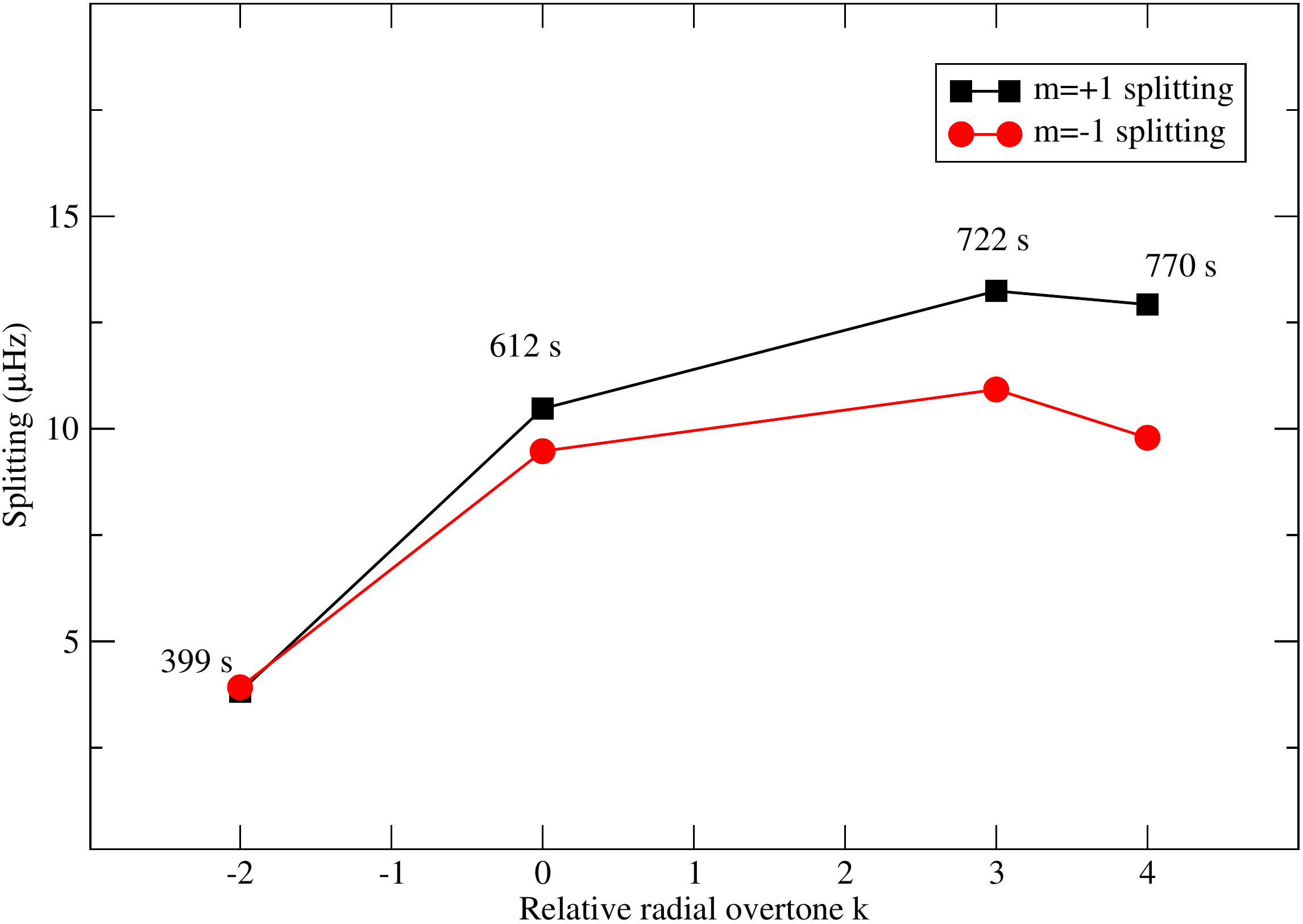}
}
\caption{Average multiplet structure associated with the large
  groupings at 770 s (1302 \muHz), 722 s (1387 \muHz), 612 s (1633 \muHz),
and 399 s (1887 \muHz). The $m= +1$ (prograde) splitting is
  always larger than the $m=-1$ splitting, and the asymmetry increases
  with increasing $k$. 
  \label{group}}
\end{figure}

We turn again to the combined observations from 2004-2008.  
The four large groupings at 768 s (1302 \muHz), 721 s (1387 \muHz), 
682 s (1466 \muHz), and 612 s (1633 \muHz) in Figure~\ref{periodplot2} 
all show indications of multiplet structure within their groups. 
We extracted an average multiplet for each of the four large groupings in
Figure~\ref{periodplot2} by calculating a simple average frequency value 
for the $m=1$, $m=0$, and $m=+1$ components.  The results are presented
in Figure~\ref{group}. In each case, the prograde ($m=+1$)
mode splitting is larger than the $m=-1$ mode, and both the splittings and 
the asymmetries increase with increasing radial node $k$.

The qualitative behavior of the low-$k$ and high-$k$ modes can be
explained in terms of a very general model. g-modes are standing waves
of buoyancy in a spherical cavity, and can be thought of as
superpositions of traveling waves bouncing back and forth between an
inner and outer turning point.  In general, low-$k$ (shorter period)
modes have deeper outer turning points while high-$k$ (longer period)
modes have turning points closer to the stellar surface, meaning that
these modes sample the outer regions of the star more than do low-$k$
modes.  Figure~\ref{group} shows that the unknown process affecting
EC14012-1446's multiplet structure acts more dramatically on the
high-$k$ modes, arguing that the structural perturbation must be in
the outer layers. A surface magnetic field and/or the convection zone
are obvious candidates. In addition, we should consider both radial
and latitudinal differential rotation.  Work is underway to improve
our understanding of multiplet structure in pulsating white dwarfs
\citep{Dalessio12}.

\begin{figure*}[t]
  \centering{
\includegraphics[width=0.9\textwidth]{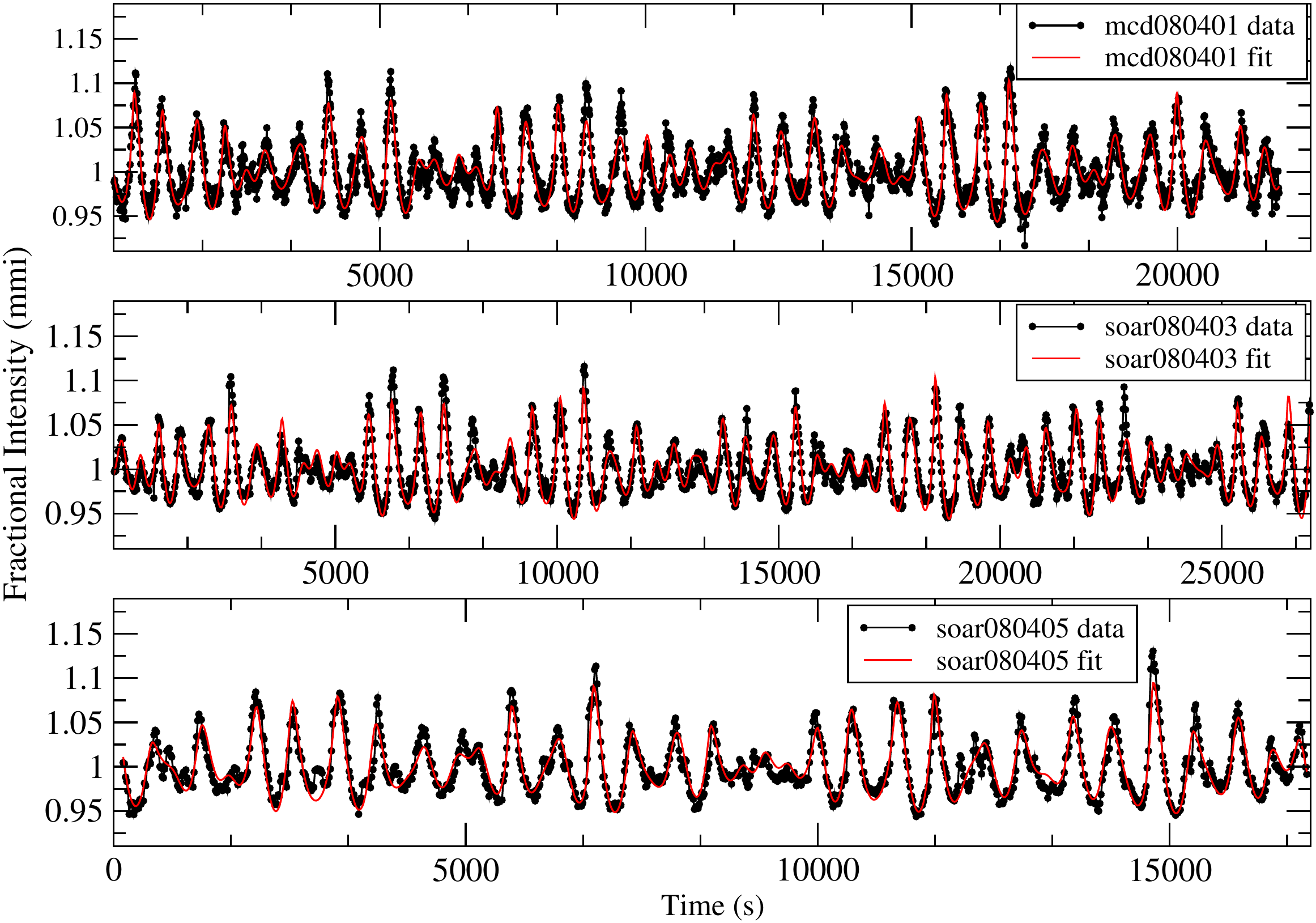}
}
\caption{Simultaneous fit of the periods of 19 modes (solid line) to
  the light curves from mcdo080401, soar080403, and soar080405 (filled
  circles).  Note the change in x axis for each plot. 
  \label{curvefits} }
\end{figure*}

Asteroseismology of EC14012-1446's distribution of excited frequencies
shows that the observed pulsations are dominated by a series of $l$=1
modes. If $l$=2 modes are present, the modes are low amplitude and
will not greatly affect the nonlinear light curve fitting process. For
the 1887, 1663, and 2504~\muHz\ multiplets, we are able to constrain
the $m$ identifications as well. The combination of these results with
the combination amplitude analysis (Table~\ref{tab:freqid}) gives us a
strong foundation on which to proceed to nonlinear light curve
fitting.

\section{Nonlinear Analysis}\label{sec:nonlinear}

\citet{Montgomery05a} and \citet{Montgomery10} give a detailed
description of fitting observed non-sinusoidal light curves of white
dwarfs to extract the time-averaged thermal response timescale of the
convection zone. To summarize, the aspect of the convection zone
sampled by the pulsations is the thermal response timescale,
$\rm{\tau_c}$, which is directly related to the convection zone's mass
and depth, and therefore its heat capacity. A DAV such as EC14012-1446
will experience local temperature excursions of $\pm 250$~K. In
response, the local mass and depth of the convection zone will vary
throughout a pulsation cycle.

MLT predicts that $\rm{\tau_c}$ should scale as
\begin{equation}
{\tau_c\approx\tau_0\left(T_{\rm eff}/T_{\rm eff,0}\right)^{-N} }
\end{equation}
where \teff\ is the instantaneous effective temperature, \teffo\ is
the equilibrium \teff, $\rm{\tau_0}$ is the time-averaged convective
timescale, and $N$ describes the sensitivity of $\rm{\tau_c}$ to changes
in \teff. For DAs, MLT predicts $N \approx 90$ \citep{Montgomery05a, 
Wu01, Brickhill92a}. The convection zone's mass, depth, and heat
capacity are therefore extraordinarily temperature sensitive, and can
vary by a factor of $\approx 10$ throughout a pulsation cycle. This
modulation of depth and heat capacity is the source of the large
nonlinearities in white dwarf light curves. We focus on these observed
nonlinearities to determine the time-averaged convective timescale
$\rm{\tau_0}$, the temperature sensitivity parameter $N$, and the
inclination of the pulsation axis to the line of sight $\rm{\theta}$.

Our analysis follows the approach of
\citet{Montgomery10}. EC14012-1446 is a multiperiodic pulsator, and
because of nonlinear effects, a pulse shape obtained by folding its
light curve at a period of interest is not equivalent to a pulse shape
obtained in the absence of additional frequencies
\citep{Montgomery07a}. Therefore, we use the accurate frequencies and
($l,m$) identifications obtained for the 2008 WET campaign to
calculate point by point nonlinear light curve fits to light curves
obtained during XCOV26. High signal-to-noise light curves
(S/N$\approx$1000) are vital, since we are interested in the nonlinear
portion of the data, which is smaller than the linear component. We
chose the SOAR 4m light curves, and the two longest runs from the
McDonald 2.1m (selected runs are marked with ``*'' in
Table~\ref{tab:journal}. These 6 runs span three days, a timebase long
enough to constrain the phases of the closely spaced frequencies, yet
short enough to avoid possible implications of amplitude modulation
found in our frequency analysis (see Section~\ref{sec:stab}). The
dominant frequencies can be considered to be stable over a 3 day
timescale. Using model atmosphere tables provided by D. Koester, we
calculate the conversion from bolometric to the observed bandpass as
described in \citet{Montgomery10}, assuming the following parameters
for this star: \teffo=$11768\pm23$~K and log g=$8.08\pm0.008$ in cgs
units \citep{Koester09}.

We began the fitting process by including the frequencies in
Table~\ref{tab:freqid} with firm ($l,m$) identifications.  We then
included the additional frequencies with strong ($l,m$)
constraints. It is prudent to point out that this fitting process is
{\it nonlinear}. Adding additional frequencies based on criteria such
as amplitude is not necessarily the best procedure.  We experimented
extensively by computing numerous fits encompassing a wide range of
($l,m$) identifications for the lower amplitude frequencies. The
values of ${\tau_0}$ for all fits range from 99--230~s, indicating
that ${\tau_0}$ is not strongly dependent on these identifications.

Our investigation does reveal that the temperature parameter $N$ can
be highly sensitive to the input $m$ identification for large
amplitude modes. Numerous attempts to fit EC14012-1446's light curve
while assigning the 1633.907 \muHz\ mode an ($l,m$) identification of
(1,0) as derived from our combination analysis (see
Section~\ref{sec:comfreq}) yielded ${\tau_0}$=171 s,
${\theta}=28^\circ$, and $N=39$. While the values of ${\tau_0}$ and
${\theta}$ are reasonable values based on MLT, the value of $N$ implies a temperature
sensitivity that is far below predictions of MLT \citep{Montgomery10};
this is not a physically relevant fit. We also experimented by
assigning a spherical index $l$=2 to this mode, resulting in the
following fit parameters: ${\tau_0}=228$~s, $N=26$,
${\theta}=15^\circ$. Again, the value of $N$ obtained remains too
low. \citet{Ising01} warn that a perturbation analysis of combination
amplitudes \citep{Wu01} may have difficulties for photometric
variations of high amplitude, and this mode has the largest amplitude
in the data set. In addition, the asteroseismology analysis of the
1633~\muHz\ multiplet structure cannot solidly constrain the $m$ value
for this mode (see Section~\ref{sec:astero}).  Following this
reasoning, we experimented by assigning the 1633.907~\muHz\ mode an
($l,m$) value of (1,1). The resulting fit finds $N=85$, in much better
agreement with MLT predictions \citep{Wu01,Montgomery05a}.

The final simultaneous nonlinear fit to the 6 high signal-to-noise light curves
includes the frequencies and ($l,m$) identifications given in
Table~\ref{tab:freqid}. This fit produced the following parameters:
${\tau_0}=99.4\pm12$ s, $N=85\pm6$, ${\theta}=32.9\pm3.2^\circ$.
Figure~\ref{curvefits} shows the ability to reproduce the essential
features of the light curves.

\section{Discussion}

Convective energy transport in stellar environments is typically
modeled using MLT. A particular version by \citet{Bohm71}, denoted as
ML2, includes reduced horizontal energy loss relative to the
formulation of \citet[ML1,][]{Bohm-Vitense58}, increasing the overall
convective efficiency. ML2 has been the standard convection model
adopted for stellar atmosphere fits of white dwarfs for the past 20
years, with $\alpha=0.6$ the preferred value for the mixing length
\citep{Bergeron95}.  \citet{Tremblay2010} recently re-calibrated the
assumed convective efficiency for white dwarf models, using model
spectra incorporating an improved treatment of Stark broadening
\citep{Tremblay09}.  \citet{Tremblay2010} fit an improved set of
Hubble Space Telescope and International Ultraviolet Explorer
ultraviolet (UV) and near UV spectra, varying ${\alpha}$ until
reaching agreement between the optical and UV temperatures.  They find
the best internal consistency between optical and UV effective
temperatures and log g measurements using ML2 with ${\alpha}$=0.8.
This is a much more efficient version of MLT than found by
\citet{Bergeron95}, but is in closer agreement with that required by
nonadiabatic models to fit the observed blue edge of the DA
instability strip \citep{Fontaine08}. This convective parameterization
is becoming the standard for DA model atmospheres \citep{Freytag12,
  Kilic12}.

\begin{figure}[t]
  \centering{
\includegraphics[width=1.00\columnwidth]{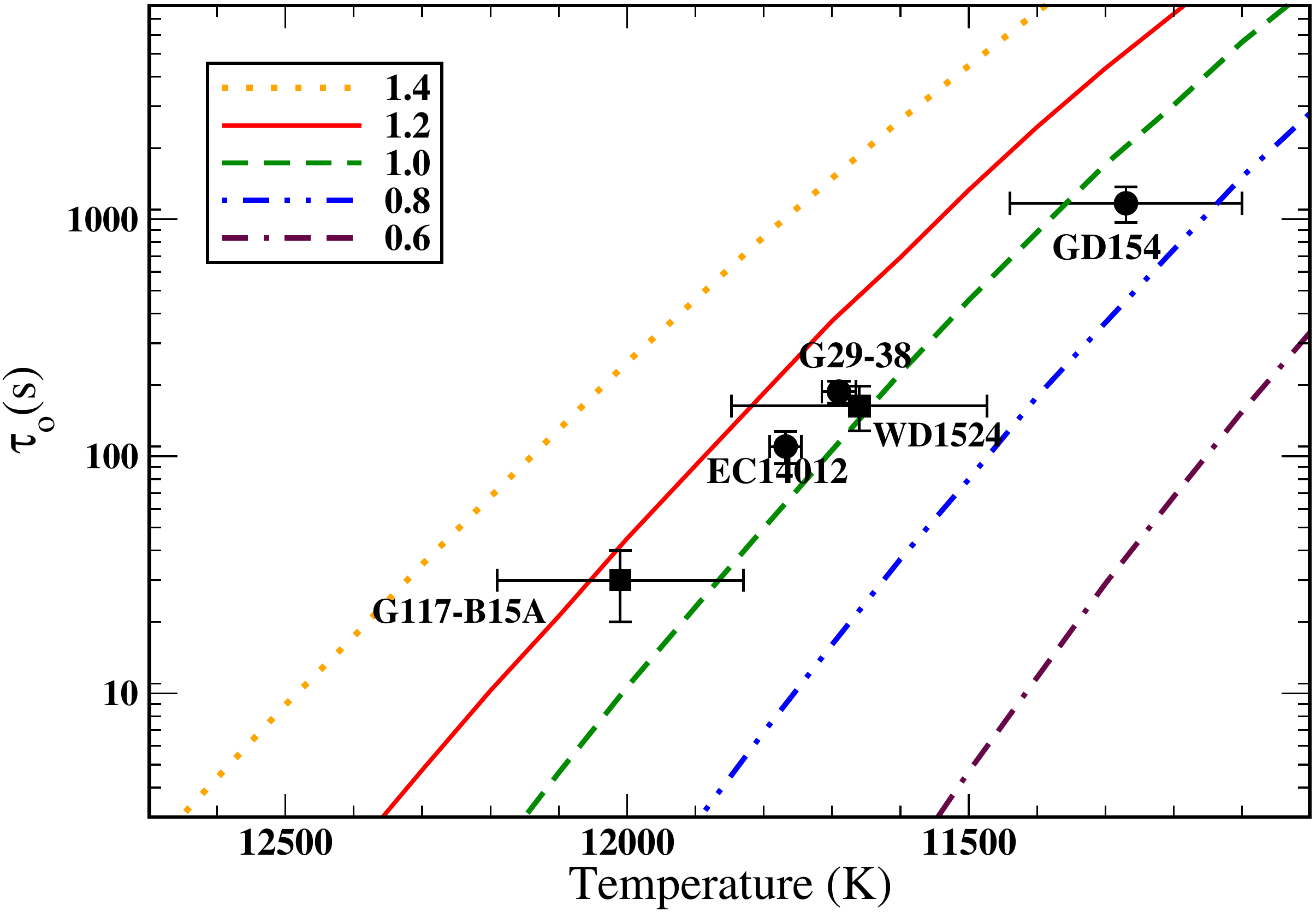}
}
\caption{Comparison of EC14012-1446's derived convective parameters
  with values expected from ML2 convection.  The additional labeled
  points are individual objects taken from \citet{Montgomery05a},
  \citet{Montgomery10c}, and \citet{Provencal11}. The curves represent
  ML2 theoretical calculations of the thermal response time ${\tau_0}$ 
for various values of the mixing length $\alpha$ (log g=8.0). 
\label{dastars}}
\end{figure}

One of our goals is to produce an empirical map of the time-averaged
convective timescale ${\tau_0}$ at the base of the convection zone as
a function of \teff\ and $\log g$ for a population spanning the DAV
instability strip ($\approx 11,100$--$12,200$~K) and compare this with
MLT predictions of ${\tau_0}$. As an individual DA white dwarf cools
through the instability strip, the base of its surface hydrogen
convection zone will deepen, increasing its total mass and the value
of ${\tau_0}$. MLT makes specific predictions for the behavior of
${\tau_0}$ given different choices of input parameters.  For instance,
Figure~\ref{dastars} shows that ML2 with ${\alpha}$=1.0 predicts
${\tau_0}$ values ranging from $\sim 1$~s at the blue edge of the
instability strip to $\sim 6000$~s at the red edge.

\begin{deluxetable*}{rrrrrrrr} 
\tablecaption{Stellar Parameters}
\tablehead{
\colhead{Star} & \colhead{\teff(K)} & \colhead{log g} &
\colhead{Period (s)} & \colhead{$\tau_0$(s)} &\colhead{$\rm{P_{max}}$}& \colhead{$\theta$} &
\colhead {$N$} \\
}
\startdata
G117-B15A & $12010\pm180$ & 8.33 & 215.2 &$30\pm\phantom{1}10$ &
$188\pm\phantom{10}63$ & --- & ---\\
EC14012-1446 & $11768\pm\phantom{1}23$ & 8.11 & 612.3 &
$99\pm\phantom{1}12$ &
$624\pm\phantom{10} 75$ &$33\pm\phantom{1}3$ & $85\pm\phantom{1}6$  \\
G29-38 & $11690\pm120$ & 8.11 & 615.2& $187\pm\phantom{1}20$ &
$1177\pm\phantom{1}126$
&$66\pm\phantom{1}3$ & $95\pm\phantom{1}8$  \\
WDJ1524-0030 & $11660\pm180$ & 8.06 & 697.3 & $163\pm\phantom{1}35$
&$1024\pm\phantom{1}220$ &$58\pm10$ & $95\pm15$  \\
GD154 & $11270\pm170$ & 7.95& 1185.9 & $1169\pm200$ & $7345\pm1257$ &
$10\pm10$ & $127\pm25$ 
\enddata
\label{tab:stars}
\tablecomments{\teff\ measurements are from \citet{Koester09},
  \citet{KH01}, \citet{Koester00}.  Given pulsation periods are for
  the largest amplitude frequency. Entries marked with ``---'' did not
  have unique values determined by the fits.  }

\end{deluxetable*}

In Figure~\ref{dastars} we also plot current determinations of
${\tau_0}$ versus \teff\ for five DAVs, including our solution for
EC14012-1446; the derived stellar parameters are listed in
Table~\ref{tab:stars}. To ensure a uniform treatment of all the stars
in our sample, we have used \teff\ determinations that do not include
the recent updates to the line profile calculations by
\citet{Tremblay09}; rather, the plotted effective temperatures and
horizontal error bars are based on earlier spectroscopic fits
employing ML2 \citep{Koester09, KH01, Koester00}. Overlaid on this
figure are the ${\tau_0}$ predictions of ML2 convection for various
values of the mixing length parameter ${\alpha}$. Our current results
do indicate an increase in ${\tau_0}$ (and hence an increase in depth
and mass of the convection zone) with decreasing temperature, and are
marginally consistent with ${\alpha}=1.0$. Decreasing the \teff\ error
bars, either through higher signal-to-noise spectra and/or new model
atmosphere fits, will provide more precise constraints on convection
in these stars.

Finally, we point out that convective light curve fitting extracts the
value of ${\tau_0}$ at the base of the convection zone. There is no
reason why the same value of $\alpha$ should describe both the
photosphere and the deeper convective layers, so our results do not
necessarily have to agree with the results of \cite{Tremblay2010} or
\citep{Bergeron95}. For instance, \citet{Ludwig94} use sophisticated
2-D hydrodynamic simulations to show that, while MLT is a reasonable
approximation to predict the rough photospheric temperature structure
of DA white dwarfs, the deeper layers have a higher convective
efficiency than predicted by MLT.

\begin{figure}[b]
  \vspace*{0.5em}
 \centering{
\includegraphics[width=1.0\columnwidth]{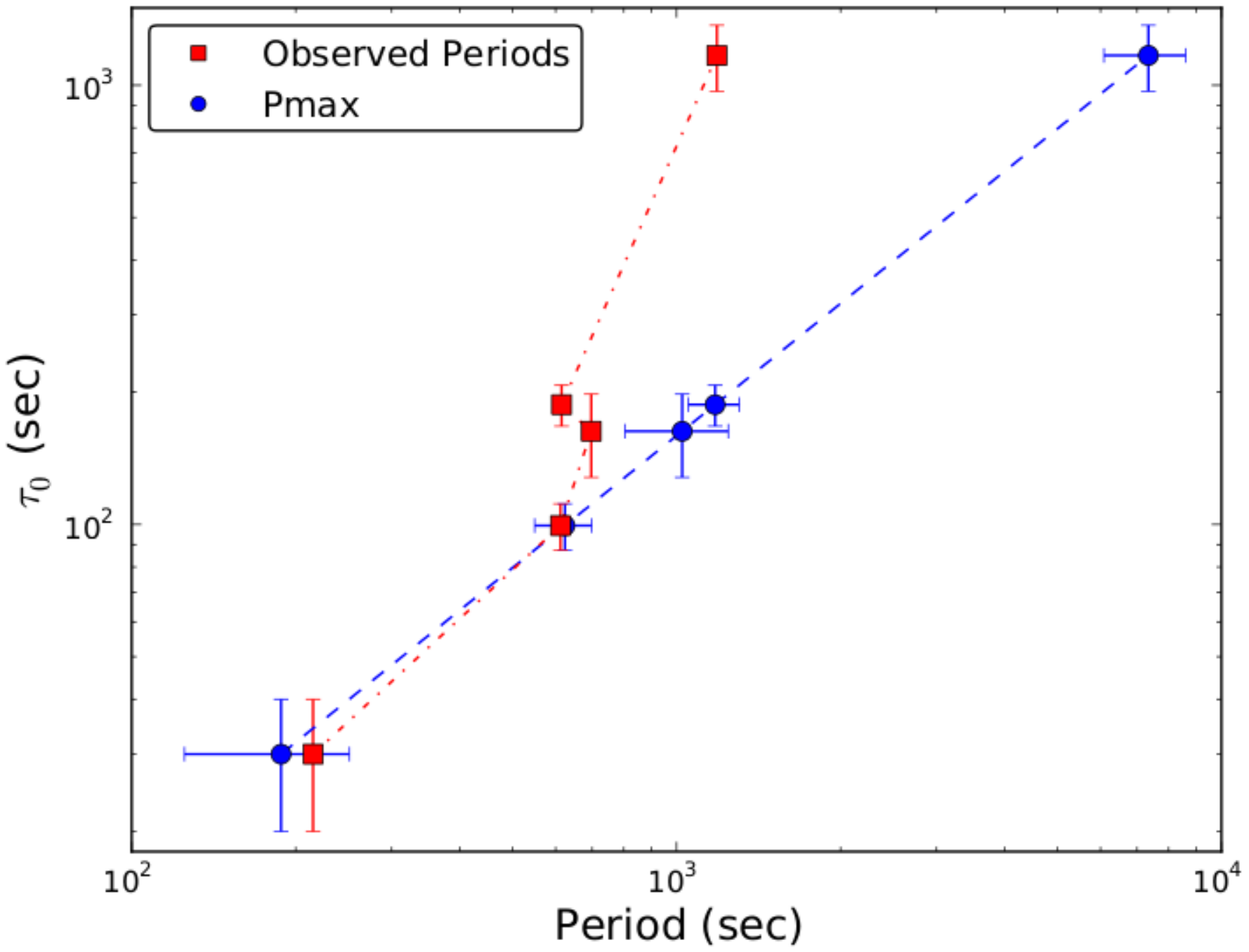}
}
\caption{Comparison of the dominant observed pulsation period in a
  star (dot-dashed curve, red squares) with the theoretical value of
  $P_{\rm{max}}$ (dashed line, blue circles) for each DAV star with a
  measured value of $\tau_0$. Cooler stars are to the left, hotter
  stars to the right.  While simple theory predicts that these curves
  should be very similar, we find a significant departure for cooler
  stars. 
\label{pmax}
}
\end{figure}

Our treatment of the nonlinearities in white dwarf light curves is
based on the larger picture of how a surface convection zone leads to
driving in these stars. In particular, \citet{Brickhill91a} and
\citet{Goldreich99a} demonstrate that excitation of g-mode pulsations
should occur when the convective driving exceeds the radiative
damping, and that this condition is given by $\omega\tau_0 \gtrsim 1$.
In terms of period, this relation states that g-modes should be driven
when $P \lesssim P_{\rm max}$, where $P_{\rm max} \equiv 2\, \pi\,
\tau_0$ and $P$ is the period of a given mode. In Figure~\ref{pmax} we
compare the dominant oscillation period in each of the DAVs we have
fit with their theoretical value of $P_{\rm max}$, as calculated from
each star's value of $\tau_0$.  The agreement is good for $\tau_0
\lesssim 100$~s, but the values diverge for larger values of
$\tau_0$. Given that $\tau_0$ is related to a star's \teff\ (see
Figure~\ref{dastars}), this says that the agreement is good from the
blue edge to near the middle of the instability strip, but that from
this point to the red edge another effect is operating that prevents
the dominant mode periods from increasing as rapidly as $P_{\rm max}$.

It is natural to assume that the cause of this behavior is related to
whatever unknown phenomenon produces the red edge of the instability
strip. \citet{Wu01a} point out that a deepening convection zone
attenuates the flux variations entering at its base, so that for deep
enough convection zones the surface amplitudes will be below detection
limits. However, this is an amplitude effect and does not predict that
the dominant periods should be different from $P_{\rm
  max}$. \citet{Brickhill91a}, \citet{Goldreich99a}, and \citet{Wu01a}
also discuss the importance of turbulent damping in a narrow shear
layer at the base of the convection zone; damping from this region may
be more significant for cooler stars. Another possibility is that the
maximum period for which complete surface reflection of g-modes can
occur is much lower than expected; typical estimates place this number
on the order of $\sim 5000$~s for $l=1$ modes and $\sim 3000$~s for
$l=2$ modes \citep{Hansen85}, although it is possible that improved
treatment of the surface layers could alter these estimates. At any
rate, this phenomenon is likely related to the deepening of the
convection zone toward the red edge; we are exploring other phenomena
that could lead to this behavior.

\section{Conclusions}

Until the advent of convective light curve fitting, our Sun was the
only star with empirical constraints on its convection zone depth.
Determinations of the time averaged convective timescale $\tau_0$ and
the temperature sensitivity parameter $N$ can now potentially be
obtained for any white dwarf pulsator of moderate amplitude
\citep{Montgomery10}; approximately two-thirds of all white dwarf
pulsators show significant nonlinearities in their light curves.  Our
ultimate goal is to map $\tau_0$ as a function of \teffo\ and $\log g$
for a population spanning the instability strips of both the DAV and
DBV white dwarfs. Such a map will provide important empirical constraints on
convection for white dwarfs and eventually other types of pulsating
stars.

We have taken the first steps in this direction with our investigation
of EC14012-1446. XCOV26 produced 308.3 h of data, and our
analysis has identified 19 independent frequencies distributed in 14
multiplets.  Combined with archival observations, we have
identified a series of $l$=1 modes with an average period spacing of
41 seconds.  EC14012-1446 is now one of the few DAVs with over a dozen
identified modes in its pulsation spectrum. The large number of modes
means that asteroseismology can be used to provide constraints on its
interior structure.

Future goals focus on the determination of $\tau_0$ and $N$ for
additional white dwarf pulsators. An increased sample size will
improve our empirical map of $\tau_0$ as a function of \teff\ and also
allow us to further explore the observed behavior of $P$ and $P_{\rm
  max}$ as a function of $\tau_0$. Our goals also include the
reduction of errors associated with spectroscopic
temperatures. Convective light curve fitting demands uniform treatment
of convection between the spectroscopic temperatures we choose and the
temperatures we use to calculate the light curve fits. Since
WD1524-0030 was not included in the new temperatures of
\citet{Gianninas2011} that incorporate recent updates to line profile
calculations, we use a consistent set of older published temperatures
for the objects presented in this work. Incorporating the new
temperatures will be a first step towards reducing the spectroscopic
errors.  We must recalculate our existing nonlinear light curve fits
using these updated \teff\ and $\log g$ determinations. On average,
the temperatures of \citet{Gianninas2011} increase over older
published values by $\approx 500$~K. Preliminary work indicates that
the empirical value of $\tau_0$ will not change significantly with
such an increase.  However, the theoretical MLT predictions for
$\tau_0$ \emph{will} change; larger values of $\alpha$ will be needed
to keep $\tau_0$ and the convection zone depth the same for these
higher values of \teff.




\section{Acknowledgments}

The Delaware Asteroseismic Research Association is grateful for the
support of the Crystal Trust Foundation and Mt.\ Cuba
Observatory. DARC also acknowledges the support of the University of
Delaware, through their participation in the SMARTS consortium. MHM
gratefully acknowledges the support of the NSF under grant AST-0909107
and the Norman Hackerman Advanced Research Program under grant 003658-
0252-2009.  SLK acknowledges partial support by the KASI
(Korea Astronomy and Space Science Institute) grant 2012-1-410-02. This paper 
uses observations made at the South African Astronomical Observatory (SAAO).
This work is further supported by the Austrian Fonds zur
F\"orderung der wissenschaftlichen Forschung under grant P18339-N08.
We would like to thank the various Telescope Allocation Committees for
the awards of telescope time.



{\it Facilities:} \facility{MCAO:0.6m ()}, \facility{McD:2.1m ()},
\facility{KPNO:2.1m ()}, 
\facility{UH:0.6m ()}, \facility{BOAO:1.8m ()}, \facility{Lulin:1.8m ()},
\facility{Beijing:2.16m ()}, \facility{Maidanek:1.0m ()}, \facility{Peak Terskol}

\bibliography{index_f}

\end{document}